\documentclass[times,final]{elsarticle}

\usepackage{jcomp}
\usepackage{framed,multirow}

\usepackage[utf8]{inputenc} 				
\usepackage[main=english, ngerman]{babel}	
\usepackage[T1]{fontenc} 					
\usepackage{textcomp} 					
\usepackage{amsmath}					
\usepackage{amssymb} 					
\usepackage{cases}                      
\usepackage[hidelinks]{hyperref}		
\usepackage[noabbrev]{cleveref}			
\usepackage[final]{microtype}			
\usepackage{csquotes}					
\usepackage{graphicx}                   
\usepackage{grffile}					
\usepackage{subcaption}					
\usepackage{lmodern}		            
\usepackage{multirow}                   
\usepackage{array}						
\usepackage{makecell}					
\usepackage{booktabs}					
\usepackage{placeins}					
\usepackage{xcolor}						
\usepackage{setspace}
\usepackage{lineno}

\usepackage{rotating}
\usepackage{multicol}
\usepackage{etoolbox}  
\usepackage{enumitem}  

\usepackage{algorithmic}
\usepackage[ruled, linesnumbered]{algorithm2e}
\SetAlFnt{\small}
\SetAlCapFnt{\small}
\SetAlCapNameFnt{\small}
\DontPrintSemicolon 

\usepackage{tikz}
\usetikzlibrary{decorations.pathreplacing, angles, quotes, arrows, positioning, backgrounds, shapes, decorations.fractals, spy}

\usepackage{pgfplots}
\usepgfplotslibrary{groupplots}
\usetikzlibrary{positioning,calc,decorations.pathreplacing,shapes,arrows, backgrounds}
\usetikzlibrary{decorations.markings}
\usetikzlibrary{arrows.meta}
\usetikzlibrary{fadings}
\tikzfading 
[
name=fade out,
inner color=transparent!0,
outer color=transparent!100
]
\pgfplotsset{compat=1.9}
\pgfplotsset{every tick label/.append style={font=\footnotesize}}
\pgfplotsset{every axis label/.append style={font=\footnotesize}}
\pgfplotsset{every axis legend/.append style={font=\footnotesize}}

\usepackage{txfonts}

\usepackage{mathtools}

\DeclarePairedDelimiter\abs{\lvert}{\rvert}%

\usepackage{siunitx}
\usepackage{nicefrac}

\newcommand{\bu}{\boldsymbol{u}}

\newcommand{\Rey}{\text{Re}}
\newcommand{\Ca}{\text{Ca}}
\newcommand{\Ma}{\text{Ma}}

\newcommand{\pystencils}{{\em pystencils}}
\newcommand{\lbmpy}{{{\em lbmpy}}}
\newcommand{\walberla}{\textsc{waLBerla}}
\newcommand{\sympy}{{{\em SymPy}}}

\usepackage{url}
\usepackage{xcolor}
\definecolor{lssblue}{rgb}{0.121,0.231,0.4}
\definecolor{lssred}{rgb}{0.76,0.3,0.19}

\definecolor{color1}{rgb}{0.00392156862745098, 0.45098039215686275, 0.6980392156862745}
\definecolor{color2}{rgb}{0.8705882352941177, 0.5607843137254902, 0.0196078431372549}
\definecolor{color3}{rgb}{0.00784313725490196, 0.6196078431372549, 0.45098039215686275}
\definecolor{color4}{rgb}{0.8352941176470589, 0.3686274509803922, 0.0}

\usepackage{acronym}
\acrodef{LB}[LB]{lattice Boltzmann}
\acrodef{LBM}[LBM]{lattice Boltzmann method}
\acrodef{PF}[PF]{phase-field}
\acrodef{RK}[RK]{Runge-Kutta}
\acrodef{RK2}[RK2]{second-order Runge-Kutta}
\acrodef{RK4}[RK4]{fourth-order Runge-Kutta}
\acrodef{WMRT}[WMRT]{weighted multiple-relaxation-time}
\acrodef{PDF}[PDF]{probability distribution function}

\journal{Journal of Computational Physics}

\begin{document}

\verso{M.\ Holzer \textit{et al.}}
	
\begin{frontmatter}
	
\title{Development of a central-moment phase-field lattice Boltzmann model for thermocapillary flows: Droplet capture and computational performance}

\author[1,2]{Markus \snm{Holzer}\corref{cor1}}
\ead{markus.holzer@fau.de}
\cortext[cor1]{Corresponding author}

\author[3]{Travis \snm{Mitchell}}

\author[3]{Christopher R. \snm{Leonardi}}

\author[1,2]{Ulrich \snm{Rüde}}

\address[1]{Chair for System Simulation, Friedrich-Alexander-Universität Erlangen-Nürnberg, Cauerstraße 11, 91058 Erlangen, Germany}

\address[2]{CERFACS, 42 Avenue Gaspard Coriolis, 31057 Toulouse Cedex 1, France}

\address[3]{School of Mechanical and Mining Engineering, The University of Queensland, St Lucia, QLD 4072, Australia}

\received{...}
\finalform{...}
\accepted{...}
\availableonline{...}
\communicated{...}


\begin{abstract}
\begin{spacing}{1.0}
This study develops a computationally efficient phase-field \ac{LB} model with the capability to simulate thermocapillary flows. The model was implemented into the \textit{open-source} simulation framework, \walberla{}, and extended to conduct the collision stage using central moments.
The multiphase model was coupled with both a passive-scalar thermal \ac{LB}, and a \ac{RK} solution to the energy equation in order to resolve temperature-dependent surface tension phenomena.
Various lattice stencils (D3Q7, D3Q15, D3Q19, D3Q27) were tested for the passive-scalar \ac{LB} and both the second- and fourth-order \ac{RK} methods were investigated.
There was no significant difference observed in the accuracy of the \ac{LB} or \ac{RK} schemes. The passive scalar D3Q7 \ac{LB} discretisation tended to provide computational benefits, while the second order \ac{RK} scheme is superior in memory usage. This paper makes contributions relating to the modelling of thermocapillary flows and to understanding the behaviour of droplet capture with thermal sources analogous to thermal tweezers.
Four primary contributions to the literature are identified.
First, a new 3D thermocapillary, central-moment phase-field \ac{LB} model is presented and implemented in the open-source software, \walberla.
Second, the accuracy and computational performance of various techniques to resolve the energy equation for multiphase, incompressible fluids is investigated.
Third, the dynamic droplet transport behaviour in the presence of thermal sources is studied and insight is provided on the potential ability to manipulate droplets based on local domain heating.
Finally, a concise analysis of the computational performance together with near-perfect scaling results on NVIDIA and AMD GPU-clusters is shown.
This research enables the detailed study of droplet manipulation and control in thermocapillary devices by providing a highly-efficient computational modelling methodology.
\end{spacing}
\end{abstract}

\end{frontmatter}


\section{Introduction}\label{sec:intro}

The temperature driven motion of bubbles and droplets, also known as thermocapillary or Marangoni convection, can be observed in nature and is important in many naturally occurring processes and numerous industrial ventures~\cite{kotz_2004,PhysRevE.105.015314,Abe2004269,ELBOUSEFI2023124049}. Applications involving microfluidic devices or reduced gravity environments progress into a regime where the interfacial forces become increasingly dominant. The ability to control these forces through applied thermal gradients represents an attractive means of manipulation. For most common two-fluid systems, the surface tension will vary with temperature leading to a shear force along the interface and resulting in dynamic behaviour. Typically the surface tension between two fluids has an inverse relation with temperature causing fluids to migrate from regions of hot to cold, acting to minimise the total surface energy~\cite{kotz_2004,PENDSE20101147}. Further, using this temperature-dependent surface tension has been explored in the literature for the design of heat transfer equipment~\cite{smith_1995}. There also exists complex (e.g. long-chain alcohols, liquid metallic alloys) fluids that show a parabolic (or non-linear) relation between surface tension and temperature~\cite{Abe2004269,shanahan_2014,ELBOUSEFI2023124049,mitchell2021computational}. This can be used to generate recalcitrant bubbles capable of migrating in the opposing direction to the flow of the continuous phase~\cite{mitchell2021computational}. 

Effective manipulation of droplets has contributed to the development of droplet-based microfluidic devices capable of rendering programmable and re-configurable operations~\cite{Baroud2006ThermocapillaryVF,sohrabi_kassir_keshavarz_moraveji_2020}. Various techniques have been applied to control and move droplets, however, researchers have found that using surface tension forces can be more effective than radiation pressure forces (i.e. optical traps/tweezers). These thermal, or local laser-heating, techniques have been applied to move droplets as small as 30\,$\mu$m and up to 1.5\,mm~\cite{kotz_2004}. With this active control of droplet transport and configurable generation, droplets represent natural candidates to facilitate microfluidic reaction vessels~\cite{baroud2007}. This study aims to provide insight into the impact of local laser-heating on droplet transport, and provide a highly-efficient numerical framework to simulate this behaviour allowing for future design of microfluidic systems.

The \ac{LBM} has seen increasing popularity in the last decade for modelling of thermocapillary driven flows~\cite{LIU20124433,PhysRevE.87.013010,liu_valocchi_zhang_kang_2014,mitchell2021computational,PhysRevE.105.015314,ELBOUSEFI2023124049}. Studies that use this approach are able to supplement experimental research in this field by providing an opportunity for complete interrogation of the domain to extract precise measurements of the local temperature and flow fields during the transport of droplets. This is challenging experimentally, and there are still many open questions around the dynamics of thermocapillary droplet migration~\cite{liu_valocchi_zhang_kang_2014}. Within the framework of thermocapillary \ac{LBM}, there are numerous approaches to introduce the thermal field and its evolution. These range between coupling to a finite-difference based Runge-Kutta integration scheme~\cite{mitchell2021computational} to employing a thermal \ac{LBM} on a lattice stencil (the number of velocities included often left to the preference of the researcher)~\cite{liu_valocchi_zhang_kang_2014}. In addition to this variation, the choice of lattice discretisation and relaxation scheme (e.g. single-relaxation, multiple-relaxation-time, central-moment-relaxation) is often taken without understanding of the influence this has on the accuracy and computational efficiency of the numerics. As a secondary aim of the present study, the impact of the relaxation model for the hydrodynamics, interfacial dynamics, and thermal evolution was studied and the accuracy of using a thermal \ac{LBM} in contrast to a finite-difference Runge-Kutta approach was compared.

In this study, a central-moment, phase-field \ac{LBM} was developed and applied to study the dynamics of droplet capture with local laser-heating. This was conducted in a three-dimensional environment to capture the inherent nature and potential lateral instability of the droplet, as well as to examine the applicability of a single-heated location in comparison with dual-heating that may be analogous to laser-tweezers. The performance and accuracy of the computational algorithm was investigated by comparing a pure \ac{LBM} implementation with a hybrid scheme which uses a finite-difference based Runge-Kutta integration scheme. Various lattice stencils were applied for the thermal \ac{LB} and both the second- and fourth-order \ac{RK} schemes were examined. These assessments were made on the simulation results obtained from a layered thermocapillary flow before the optimal configuration was applied to first study the capture of a droplet in two-dimensions. The domain was then extended to three-dimensions to investigate the breakdown in symmetry and capture the behaviour of a thermocapillary droplet on a substrate. Finally, a detailed performance analysis is conducted outlining the computational efficiency of the various implementations in \walberla{} and the scalability across large GPU-based supercomputer systems \cite{waLBerla}.

This manuscript starts by presenting the governing equations and numerical methods applied to resolve these in \cref{sec:Methods}. After presenting the solution methodology, we then discuss scalable and maintainable implementation techniques that may be useful for other researchers looking to develop their own research software or make use of the workflows described here. \cref{sec:Results} initially, provides the verification benchmark of a 2D planar heated channel is investigated to compare the use of an \ac{LB} verse \ac{RK} solver for the energy equation. The 2D benchmark is extended to a pseudo-3D simulation to then determine the sensitivity of results to the applied lattice stencil. With the verification and methodology comparison complete, we then showcase the model's capability by first studying the shear-driven motion of a droplet in 2D and its interaction with a local heat source. Comparisons are made with existing literature. To extend on the state-of-the-art, this case is extended to 3D, allowing for instability in the transverse direction to impact the droplet-heat source interaction. The potential of using two heat sources to control the droplet motion in 3D is presented as the final demonstration of the model. With its potential use highlighted, we conclude the manuscript with a detailed analysis of the computational performance of the algorithms and implementations in \walberla{}.

\section{Methodology} \label{sec:Methods}
This study uses the \ac{LBM} to resolve the incompressible Navier-Stokes behaviour coupled with a conservative Allen-Cahn equation to capture the dynamics of the liquid-gas interfaces present in the system and an energy equation to model the heat transport. The model builds on the work of Mitchell~et~al.~\cite{mitchell2021computational} by extending the \ac{LB} relaxation kernel using central moments~\cite{gruszczynski2020cascaded} and compares various common techniques for solving the energy equation.

\subsection{ Governing equations } \label{subsec:govEqs}
The continuity and Navier-Stokes equations that describe the behaviour of an incompressible multiphase flow are commonly written as,

\begin{align}
	\frac{\partial \rho}{\partial t} + \nabla\cdot\rho\boldsymbol{u} &= 0, \label{eqn-continuity}\\
	\label{eqn-momentum}
	\rho \left(\frac{\partial \boldsymbol{u}}{\partial t} + \boldsymbol{u} \cdot \nabla \boldsymbol{u} \right) &= -\nabla p + \nabla \cdot \left( \mu \left[\nabla \boldsymbol{u} + (\nabla \boldsymbol{u})^T\right]\right)+ \boldsymbol{F}_s + \boldsymbol{F}_b, \\
	\label{eqn-phasefield}
	\frac{\partial \phi}{\partial t} + \boldsymbol{\nabla} \cdot (\phi \boldsymbol{u} ) &= \boldsymbol{\nabla} \cdot M \left( \boldsymbol{\nabla}\phi - \frac{1-(\phi - \phi_0)}{\xi}\boldsymbol{n}\right), \\
	\frac{\partial T}{\partial t} &= - \boldsymbol{u} \cdot \nabla T + \frac{1}{\rho c_\mathrm{p}} \left( \nabla \kappa \cdot \nabla T + \kappa \nabla^2 T \right) + q_\mathrm{T}. \label{eqn-temperature}
\end{align}

\noindent The first of these represents the continuity equation with $\rho \equiv \rho(\boldsymbol{x}, t)$ representing the local density of the fluids, $\boldsymbol{u}\equiv \boldsymbol{u}(\boldsymbol{x}, t)$ the velocity, and $t$ the time. \cref{eqn-momentum} represents the momentum equation formulated in an incompressible form with the hydrodynamic pressure, $p \equiv p(\boldsymbol{x}, t)$. The dynamic viscosity is described by $\mu \in \mathbb{R^{+}}$ and the surface tension and body force are described by $\boldsymbol{F}_s  \equiv \boldsymbol{F}_s(\boldsymbol{x}, t)$ and $ \boldsymbol{F}_b  \equiv \boldsymbol{F}_b(\boldsymbol{x}, t)$, respectively. \cref{eqn-continuity} and \cref{eqn-momentum} form the Navier-Stokes equations for incompressible multiphase flows. The Allen--Cahn equation used to track the interface is shown in \cref{eqn-phasefield}. Here, the mobility is denoted by $M \in \mathbb{R^{+}}$, the interface width by $\xi \in \mathbb{N^{+}}$, and $\boldsymbol{n}\equiv\boldsymbol{n}(\boldsymbol{x}, t)=\boldsymbol{\nabla}\phi/|\boldsymbol{\nabla}\phi|$ is the unit vector normal to the liquid--gas interface.
The principle behind phase-field models is to allocate an additional scalar field for the phase indicator parameter, $\phi \equiv \phi(\boldsymbol{x}, t)\in [\phi_{\text{L}}, \phi_{\text{H}}]$.
This phase indicator represents solely the fluid with higher density when $\phi_H=1$ and solely the lower density fluid when $\phi_{\text{L}}=0$. Additionally, the evolution of a temperature field $T \equiv T(\boldsymbol{x}, t)$ is described by \cref{eqn-temperature}. This formulation can be applied for systems where the heat dissipation and compression work done by the pressure are negliable. The temperature equation is defined with the thermal conductivity $\kappa \in \mathbb{R^{+}}$ and the heat capacity $c_\mathrm{p} \in \mathbb{R^{+}}$. Furthermore, a heat source density can be introduced with $q_\mathrm{T}$. \par

With the so-called \textit{velocity}-based \ac{LBM} formulation, the macroscopic equations are recovered in the form of a pressure and momentum equation \cite{Zu2013}. It can be seen that the chain rule has been applied to terms in the momentum equation which separates those recovered in the \ac{LBM} scheme and those that need to be included as forcing terms (see \cref{subsec:LBEs}). The equations are resolved are,

\begin{align}
	\frac{\partial p^*}{\partial t} + c_s^2 \nabla \cdot \bu &= 0, \\
	\frac{\partial \bu}{\partial t} + \bu \cdot \nabla \bu &= - \nabla p^*-\frac{p^*}{\rho}\nabla\rho + \nabla \cdot \left(\nu[ \nabla \bu + (\nabla \bu)^T ]\right) + \frac{\nu}{\rho}\left([ \nabla \bu + (\nabla \bu)^T ]\right)\cdot \nabla\rho + \frac{1}{\rho}\left(\mathbf{F_s + F_b}\right).
\end{align}

\noindent Here, the pressure has been modified by density as $p^*=p/\rho$, the kinematic viscosity has been used as $\nu=\mu/ \rho$, and the chain rule has been applied on to the $\nabla$ terms on the right hand side of \cref{eqn-momentum}. The incompressible, or velocity formulation, \ac{LBM} allows for large gradients in momentum to be avoided at high density ratios, thereby enhancing stability at these conditions.

The forces acting on the fluid include the body force associated with gravity and the surface tension forces resulting from the liquid-gas interface.
These are given as,

\begin{align}
	\boldsymbol{F}_b(\boldsymbol{x}, t) &= \rho(\boldsymbol{x}, t) \boldsymbol{g}, \\
	\boldsymbol{F}_s(\boldsymbol{x}, t) &= \mu_{\phi}\nabla\phi(\boldsymbol{x}, t) + \frac{3}{2} \xi \sigma_T \left(\abs{ \nabla \phi^2 } \nabla T(\boldsymbol{x}, t) - \left(\nabla T(\boldsymbol{x}, t) \cdot \nabla \phi\right) \nabla \phi \right),
\end{align}

\noindent respectively, with gravitational acceleration, $\boldsymbol{g} \in \mathbb{R}^{d}$. The surface tension force consists of the chemical potential, $\mu_{\phi} \in \mathbb{R}$ and the temperature, $T(\boldsymbol{x}, t)$. The surface tension is then interpolated with,

\begin{align}
	\sigma(T) = \sigma_{\mathrm{ref}} + \sigma_{\mathrm{T}} \left(T - T_\mathrm{ref}\right),
\end{align}

\noindent where $\sigma_{\mathrm{ref}}$ is the surface tension at the reference temperature, $T_\mathrm{ref}$, and $\sigma_T = \frac{\partial \sigma}{\partial T}$ is the rate of change of the surface tension with respect to the temperature, $T$ \cite{mitchell2021computational}.

\subsection{ Lattice Boltzmann equations for interface tracking and hydrodynamics } \label{subsec:LBEs}
Discretising the conservative Allen--Cahn equation with the LBM yields,

\begin{align}
	\label{eq:lb_interface}
	h_i (\boldsymbol{x} + \boldsymbol{c}_i \Delta t, t + \Delta t) - h_i(\boldsymbol{x},t) = \Omega_{i}^h \left[ h_i^{\text{eq}}(\phi,\boldsymbol{u}) -  h_i(\boldsymbol{x},t)\right]|_{(\boldsymbol{x}, t)},
\end{align}

\noindent where the collision operator is given by $\Omega_{i}^h(\boldsymbol{x},t) \in \mathbb{R}$, the phase-field \ac{PDF}s by $h_{i}(\boldsymbol{x},t) \in \mathbb{R}$, and the phase-field relaxation time by,
\begin{equation}\label{eq:pflbm-tau}
	\tau_{\phi}= \frac{1}{\frac{1}{2} + c_s^2 \cdot M}.
\end{equation}
Thus, the mobility of the interface defines the relaxation of the interface tracking LBM step. The equilibrium distribution is given by,
\begin{equation}
    h_i^{eq}(\phi,\boldsymbol{u}) = \phi w_i\left[1+\frac{\boldsymbol{e}_i\cdot\boldsymbol{u}}{c_s^2}+\frac{(\boldsymbol{e}_i\cdot\boldsymbol{u})^2}{2c_s^4} - \frac{\boldsymbol{u}\cdot \boldsymbol{u}}{2c_s^2}\right],
\end{equation}
where $\boldsymbol{e}_i$ refers to the $i$-th lattice direction and $w_i$ refers to the $i$-th lattice wight. An overview of different lattice stencils and their lattice weight is presented in \ref{sec:stencils}. Using this formulation of the LBM step, the zeroth-order moment,
\begin{equation}
	\label{eq:zeroth_moment_phase}
	\phi\left(\boldsymbol{x}, t\right) = \sum_i h_i\left(\boldsymbol{x}, t\right),
\end{equation}
computes $\phi\left(\boldsymbol{x}, t\right)$. The density of the fluid is obtained from this as,
\begin{equation}
	\rho(\boldsymbol{x}, t) = \rho(\phi) = \rho_{\text{L}} + (\phi(\boldsymbol{x}, t) - \phi_{\text{L}})(\rho_{\text{H}} - \rho_{\text{L}}),
\end{equation} 
which simply applies linear interpolation to the phase indicator, $\phi(\boldsymbol{x}, t)$, as suggested by Fakhari~et~al.~\cite{fakhari2017improved}.
The conservative Allen--Cahn equation is recovered by applying,
\begin{equation}
	\label{eqn-phasefieldforce}
	\boldsymbol{F}^{\phi}(\boldsymbol{x},t) = \frac{4\phi(1-\phi)}{\xi} \cdot \boldsymbol{n},
\end{equation}
in the collision space according to Guo's forcing scheme~\cite{mitchell2019GasWells, gruszczynski2020cascaded, PhysRevE.65.046308}.

The lattice Boltzmann equation (LBE) for the hydrodynamics is given by,
\begin{equation}
	g_i(\boldsymbol{x} + \boldsymbol{c}_i \Delta t, t + \Delta t) - g_i(\boldsymbol{x},t) = \Omega_{i}^g \left[ g_i^{\text{eq}}(p^*,\boldsymbol{u}) - g_i(\boldsymbol{x},t)\right]|_{(\boldsymbol{x}, t)},
\end{equation}
with collision operator, $\Omega_{i}^g(\boldsymbol{x},t) \in \mathbb{R}$, for the hydrodynamic \ac{PDF}s, $g_i(\boldsymbol{x},t) \in \mathbb{R}$, and normalised pressure, $p^{*} \equiv p^{*}(\boldsymbol{x},t) = p(\boldsymbol{x},t)/( c_s^2\rho(\boldsymbol{x},t))$. Note here that the LBE is formulated such that the zeroth-order moment recovers the normalised pressure,
\begin{equation}
	\label{eq:zeroth_moment_hydro}
	p^{*}(\boldsymbol{x},t) = \sum_i g_i(\boldsymbol{x},t).
\end{equation}
Additionally, it is important to notice that for $g_i^{\text{eq}}(p^*, \boldsymbol{u}) \in \mathbb{R}$, the incompressible formulation of the equilibrium \ac{PDF}s is used,
\begin{equation}
    g_i^{eq} = w_i\left(p^* + \left[\frac{\boldsymbol{e}_i\cdot\boldsymbol{u}}{c_s^2}+\frac{(\boldsymbol{e}_i\cdot\boldsymbol{u})^2}{2c_s^4} - \frac{\boldsymbol{u}\cdot \boldsymbol{u}}{2c_s^2}\right]\right).
\end{equation}

The forcing term to recover the Navier--Stokes equation is,
\begin{equation}
	\boldsymbol{F}(\boldsymbol{x},t) = \boldsymbol{F}_s + \boldsymbol{F}_b + \boldsymbol{F}_p + \boldsymbol{F}_{\mu},
\end{equation}
which consists of terms to recover the correct pressure gradient term, $\boldsymbol{F}_p \equiv \boldsymbol{F}_{p}(\boldsymbol{x},t) \in \mathbb{R}$, the viscous forces, $\boldsymbol{F}_{\mu} \equiv \boldsymbol{F}_{\mu}(\boldsymbol{x},t) \in \mathbb{R}$, the surface forces, $\boldsymbol{F}_{s} \equiv \boldsymbol{F}_{s}(\boldsymbol{x},t) \in \mathbb{R}$, and the body forces, $\boldsymbol{F}_{b} \equiv \boldsymbol{F}_{b}(\boldsymbol{x},t) \in \mathbb{R}$. The force vector is directly applied in the collision space according to Guo's forcing scheme~\cite{mitchell2019GasWells, gruszczynski2020cascaded, PhysRevE.65.046308}.
The pressure and viscous forces are given as,

\begin{align}
	\boldsymbol{F}_p (\boldsymbol{x},t) 	  &= -p^*c_s^2(\rho_{\mathrm{H}} - \rho_{\mathrm{L}})\boldsymbol{\nabla}\phi, \\
	\boldsymbol{F}_{\mu}(\boldsymbol{x},t)  &= \nu(\rho_{\mathrm{H}}-\rho_{\mathrm{L}})\left[\boldsymbol{\nabla}\boldsymbol{u}+(\boldsymbol{\nabla}\boldsymbol{u})^T\right]\cdot\boldsymbol{\nabla}\phi,
\end{align}

\noindent where $\rho_{\text{H}}$ and $\rho_{\text{L}} $ denote the density in the heavy and light phase, respectively~\cite{fakhari_geier_lee_2016}.
The kinematic viscosity, $\nu \equiv \nu(\boldsymbol{x},t)=\tau(\phi) c_s^2$, is computed using the linearly interpolated relaxation time,
\begin{equation}
	\tau(\boldsymbol{x},t) = \tau(\phi) = \tau_{\text{L}} + (\phi(\boldsymbol{x},t) - \phi_{\text{L}})(\tau_{\text{H}} - \tau_{\text{L}}),
\end{equation}
where $\tau_{\text{H}}$ is the relaxation time of the heavy phase and $\tau_{\text{L}}$ is the relaxation time of the light phase.
It is noted here that the deviatoric stress tensor can be obtained from moments of the non-equilibrium distribution to avoid the need for finite difference approximations in the velocity field. As a result, the only non-local parameter required during the lattice Boltzmann collision process is the phase-field parameter.

\subsection{ Numerical solver for the temperature equation } \label{subsec:solvtemp}
Many different methods have been proposed in the literature to compute the evolution of the temperature field with \cref{eqn-temperature} using an explicit solver in conjunction with the flow solver \cite{Lui2013, liu_valocchi_zhang_kang_2013, mitchell2021computational}.
However, the stiffness of this system translates into stability constraints for the time step that can be overcome by implicit time discretizations,
but which in turn require an implicit solver in each time step.
Often multigrid methods are considered as the method of choice since they can be used to construct asymptotically optimal solvers
\cite{brandt1991parabolic}. With these methods, an implicit solve is asymptotically not more expensive than an explicit time step.
Additionally, such methods can be executed parallel in time \cite{horton1995space,falgout2017multigrid}. Here, however, will restrict ourselves to methods that are structurally compatible with the explicit time stepping procedure inherited from the \ac{LB} method for the hydrodynamics.
In particular, the temperature field can also be recovered by an \ac{LB} formulation \cite{liu_valocchi_zhang_kang_2014}. As an alternative, we will consider the classical explicit fourth-order \ac{RK} discretisation in time with a standard space discretization, taken from \cite{mitchell2021computational}.\par

From this starting point, this work revisits various formulations and compares their numerical accuracy and computational performance. For both formulations the heat capacity, $c_\mathrm{p}$, and the heat conduction, $\kappa$, are interpolated from the phase-field as,

\begin{align}
	\kappa &= \kappa_\mathrm{H} + \phi \left(\kappa_\mathrm{H} - \kappa_\mathrm{L}\right)\\
	c_\mathrm{p} &= c_\mathrm{p, H} + \phi \left(c_\mathrm{p, H} -c_\mathrm{p, L}\right).
\end{align}
Thus, the formulations follow the same concept as the density or the relaxation time. The explicit \ac{RK4} discretisation takes the form,

\begin{align}
	\label{eqn-rk4}
	K_1 &= \delta_t R \left(t^n, T^n \right), \nonumber \\
	K_2 &= \delta_t R \left(t^n + \frac{1}{2} \delta_t, T^n + \frac{1}{2} K_1 \right),  \nonumber \\
	K_3 &= \delta_t R \left(t^n + \frac{1}{2} \delta_t, T^n + \frac{1}{2} K_2 \right),\\
	K_4 &= \delta_t R \left(t^n + \delta_t, T^n + K_3 \right),  \nonumber \\
	T^{n + 1} &= T^{n} + \frac{1}{6} K_1 +  \frac{1}{3} K_2 +  \frac{1}{3} K_3  + \frac{1}{6} K_4  \nonumber 
\end{align}

\noindent In contrast, a \ac{RK2} scheme is also tested in this study,

\begin{align}
	\label{eqn-rk2}
	K_1 &= \delta_t R \left(t^n, T^n \right), \\
	T^{n + 1} &= \delta_t R \left(t^n + \frac{1}{2} \delta_t, T^n + \frac{1}{2} K_1 \right) \nonumber, 
\end{align}

The \ac{LBM} formulation for the LBE can be written as,
\begin{equation}
	\label{eq:lb_thermal}
	f_i (\boldsymbol{x} + \boldsymbol{c}_i \Delta t, t + \Delta t) - f_i(\boldsymbol{x},t) = \Omega_{i}^f \left[ f_i^{\text{eq}}(T,\boldsymbol{u}) -  f_i(\boldsymbol{x},t)\right]|_{(\boldsymbol{x}, t)} + \delta_t w_i q_\mathrm{T},
\end{equation}
where $f_i$ denotes the thermal \ac{PDF}s. The formulation of the LBM discretisation is similar to \cref{eq:lb_interface} with the difference that the order parameter for the thermal solver is not the phase-field, $\phi$, but the temperature, $T$. Thus, the temperature field can be recovered using the zeroth-order moment of the thermal \ac{PDF}s,
\begin{equation}
		\label{eq:zeroth_moment_thermal}
	T(\boldsymbol{x},t) = \sum_i f_i(\boldsymbol{x},t).
\end{equation}
To recover the conservative Allen--Cahn equation, it was necessary to introduce a source term according to \cref{eqn-phasefieldforce}. For the thermal LBM formulation a source term is only necessary if a heat source, $q_\mathrm{T}$, exists in the simulation. This can be, for example, a laser source as used in this study. The relaxation rate for the thermal \ac{LBM} can be calculated as,
\begin{equation}\label{eq:thlbm-tau}
	\tau_{T}= \frac{1}{\frac{1}{2} + c_s^2 \cdot \kappa},
\end{equation}
and is related to the thermal conductivity.

\subsection{ Scalable and maintainable implementation techniques }\label{sec:code_generation}

When assessing the literature, it is clear that there exists a diverse range of collision models for the \ac{LBM}, each varying in complexity. Although advanced collision models, such as those employing \textit{cumulants}, demonstrate high precision and stability (particularly in the context of highly turbulent flows) their integration with specific applications remains limited \cite{GEIER2015507}. Notably, in scenarios like multiphase flows, the SRT or MRT methods retain widespread usage, while \textit{cumulant} methods find application only through tailored modifications or filtering strategies \cite{SITOMPUL201993}. Consequently, from a software design perspective, it becomes imperative not to exclusively focus on a single collision model, but to offer a collection of diverse models.

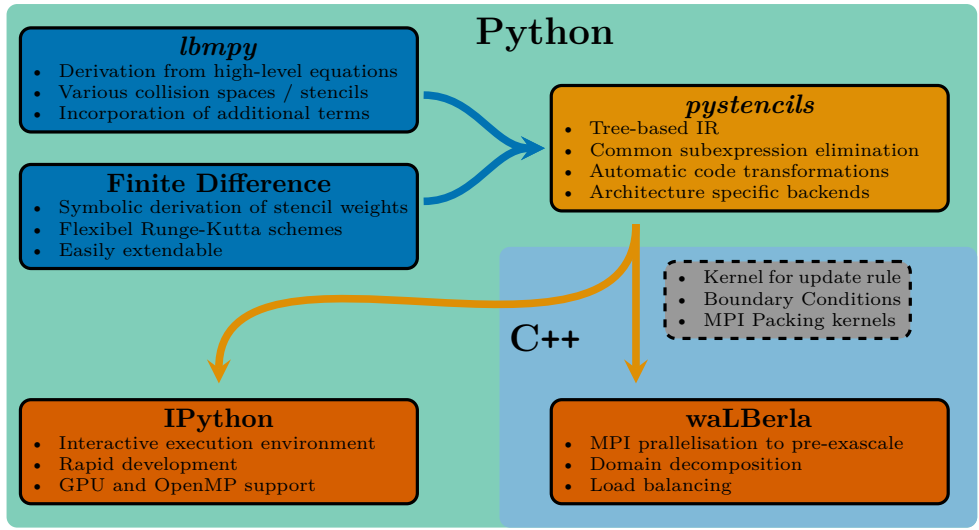
\begin{figure}[htb!]
	\centering
	\begin{tikzpicture}[]
	
	\fill[color3!50, rounded corners, dashed] (-2.8,0) rectangle (10,6.9);
	
	\fill[color1!50, rounded corners, dashed] (3.7,0) rectangle (10,3.7);
	
	\node[] at (4.3,6.5) {\textbf{\Large Python}};
	\node[] at (4.3,2.5) {\textbf{\Large C\texttt{++}}};

	\node[draw=black, very thick, rounded corners, left color=color1, right color=color1] at (0,5.9) {\parbox{5cm}{
			
			\centering
			\textbf{\lbmpy{}}
			
			{\scriptsize
				\begin{itemize}[nosep, left=0pt]
					\item Derivation from high-level equations
					\item Various collision spaces / stencils
					\item Incorporation of additional terms
			\end{itemize}}
	}};
	
	\node[draw=black, very thick, rounded corners, left color=color1, right color=color1] at (0,4.1) {\parbox{5cm}{
			
			\centering
			\textbf{Finite Difference}
			
			{\scriptsize
				\begin{itemize}[nosep, left=0pt]
					\item Symbolic derivation of stencil weights
					\item Flexibel Runge-Kutta schemes
					\item Easily extendable
			\end{itemize}}
	}};
	
	\node[draw=black, very thick, rounded corners, left color=color2, right color=color2] at (7,5) {\parbox{5cm}{
			
			\centering
			\textbf{\pystencils{}}
			
			{\scriptsize
				\begin{itemize}[nosep, left=0pt]
					\item  Tree-based IR
					\item Common subexpression elimination
					\item Automatic code transformations
					\item Architecture specific backends
			\end{itemize}}
	}};

	\node[draw=black, very thick, rounded corners, left color=color4, right color=color4] at (0,1) {\parbox{5cm}{
			
			\centering
			\textbf{\textbf{\textsc{IPython}}}
			
			{\scriptsize
				\begin{itemize}[nosep, left=0pt]
					\item Interactive execution environment
					\item Rapid development
					\item GPU and OpenMP support
			\end{itemize}}
	}};
	
	\node[draw=black, very thick, rounded corners, left color=color4, right color=color4] at (7,1) {\parbox{5cm}{
			
			\centering
			\textbf{\walberla{}}
			
			{\scriptsize
				\begin{itemize}[nosep, left=0pt]
					\item MPI prallelisation to pre-exascale
					\item Domain decomposition 
					\item Load balancing
			\end{itemize}}
	}};

	\node[draw=black, very thick, dashed, rounded corners, left color=black!40, right color=black!40] at (7.5,3) {\parbox{3cm}{

			{\scriptsize
				\begin{itemize}[nosep, left=0pt]
					\item Kernel for update rule
					\item Boundary Conditions 
					\item MPI Packing kernels
			\end{itemize}}
	}};

	\draw[-stealth, line width=1mm, color1] (2.7, 5.7)  to[out=0, in=180] (4.3, 5);
	
	\draw[-stealth, line width=1mm, color1] (2.7, 4.3) to[out=0, in=180] (4.3, 5);
	
	\draw[-stealth, line width=1mm, color2] (5.5, 4) -- (5.5, 1.9);
	
	\draw[-stealth, line width=1mm, color2] (5.5, 4) to[out=270, in=90] (0, 1.9);

\end{tikzpicture}
	\caption{Overview of the code generation methodology applied within the \walberla development framework. The two Python packages \lbmpy{} and \pystencils{} allow the flexible symbolic derivation of LBM kernels which can be executed MPI-parallel for large scale production runs within \walberla{} or as standalone application in \textsc{IPython} for rapid model development.}
	\label{fig:codegen}
\end{figure}

This introduces a distinct challenge, which is effectively tackled through the integration of meta-programming techniques within the \walberla{} framework. An overview of this approach is depicted in \cref{fig:codegen}. At the highest level, the Python package \lbmpy{} encapsulates the complete symbolic representation of the \ac{LBM}. For this, the open-source library \sympy{}, is used for symbolic manipulation of the mathematical expressions \cite{sympy}. Similarly, stencil weights for finite difference discretization can be symbolically derived as needed in \cref{eqn-momentum} and \cref{eqn-phasefield} for the calculation of the interface curvature or to solve \cref{eqn-temperature} by a coupled Runge-Kutta / finite difference method. Such a workflow permits the systematic dissection of the \ac{LBM} into its constituent parts, subsequently modularising and streamlining each step. A comprehensive depiction of this process can be found elsewhere in the literature~\cite{Hennig2023, BAUER2021101269}. In essence, the derivation of equilibrium, transformation to the collision space, computation of macroscopic properties, and incorporation of force terms or turbulence models are entirely decoupled. These stages are independently derived and simplified before being put together into a coherent update rule. From this, an assignment list emerges, which comprehensively characterises the entire LB update rule. The optimisations at each juncture enables the generation of highly specialised, problem-specific LBM compute kernels with minimal Floating Point Operations (FLOPs), all while maintaining a degree of modularity within the source code.

From the symbolic description, an Abstract Syntax Tree (AST) is constructed within the \pystencils{} Intermitted Representation (IR) where spatial access is encapsulated through \pystencils{} \emph{fields} \cite{Bauer19}. This, enables the incorporation of architecture-specific AST nodes to represent, for example, loop nests or pointer access in subsequent compute kernels. Thus, optimization strategies like OpenMP parallelization or SIMD vectorization can be seamlessly applied. Additionally, such loop nests can be entirely transformed to fit accelerators like GPUs but also constant variables can be entirely removed outside to allow pre-calculation before entering the loop-nest. Since now all data accesses within the \pystencils{} IR are defined symbolically, it is possible to fully automate the creation of kernels for packing and unpacking, which plays a pivotal role in populating communication buffers for MPI operations. Similarly, fitting boundary conditions can be derived from a consistent symbolic form.

Finally, within \pystencils{} a clear and simple interface is defined for printing the compute kernels with the C or GPU backend. The base paradigm of which is that each generated function only takes raw C-pointers for array access with their representative shape and stride information and free parameters. This simple and consistent interface makes it possible to easily integrate the kernels in existing C/C++ software structures. In the case of Python, the Python C-API can be employed. This means, additionally, we can provide a powerful interactive development environment for rapid model development by utilizing \lbmpy{}/\pystencils{} as stand-alone packages.

\section{Results}
\label{sec:Results}
This work first assessed the use of an \ac{LBM} to resolve the thermal field in comparison to using a \ac{RK2} or \ac{RK4} integration scheme by comparing steady-state results with analytical expressions for 2D layered thermocapillary fluids. Following this, the importance of the lattice stencil was investigated with the same test case, but extended in the third Cartesian direction. This allowed the thermal \ac{LBM} performance to be examined using a D3Q$q$ lattice where $q\in\{7,15,19,27\}$. From the findings of these investigations, the choice in numerical method for resolving the thermal field as well as the desired lattice for 3D simulations was determined. This was then applied to investigate thermocapillary droplet trapping using locally heated points in the flow domain. Firstly, comparison with existing literature was conducted, however, these tests are limited to 2D. To extend on the literature and generate insight on the behaviour of thermocapillary droplet capture with, for example, laser-tweezers or manipulation through other thermal sources, a 3D domain was constructed and the dynamics of a thermocapillary droplet were analysed in the presence of one- and two-heat sources. The analysis of the test sections serve to both verify and contrast the implementations before showcasing their effectiveness to generate understanding on droplet capture. Following this, the computational performance of the various implementations are detailed, both in single-GPU performance and in scaling across \num{4096} GPUs.

\subsection{Verification: 2D Planar heated channel to compare \ac{LB} and \ac{RK} schemes}
\label{sec:planarChanel2D}
This section benchmarks the coupled thermocapillary, central-moment, \ac{PF} \ac{LBM} presented in \cref{sec:Methods}. To do this, the thermocapillary-driven motion of fluid in a heated microchannel is investigated. This problem has been addressed by numerous authors \cite{pendse_esmaeeli_2010, liu_valocchi_zhang_kang_2013, mitchell2021computational} to verify thermocapillary flow models, and is used here for this purpose, as well as to compare the computational performance of various model details. The solution methodology for the energy equation is analysed with a comparison made between a passive-scalar LBM, \ac{RK2}, and \ac{RK4} integration scheme.

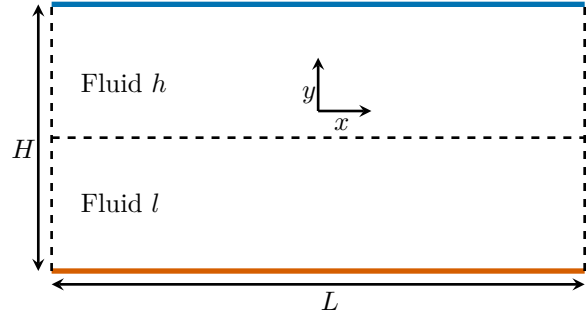
\begin{figure}[htb]
	\centering
    \begin{tikzpicture}[x=1pt, y=1pt]
	
	\draw [color4, draw opacity=1 ][line width=2]    (0,0) -- (200,0) ;
	\draw [black, draw opacity=1, dashed ][line width=1]    (0,50) -- (200,50) ;
	\draw [color1, draw opacity=1 ][line width=2]    (0,100) -- (200,100) ;
	
	\draw [black, draw opacity=1, dashed ][line width=1]    (0,0) -- (0,100) ;
	\draw [black, draw opacity=1, dashed ][line width=1]    (200,0) -- (200,100) ;
	
	\draw [draw opacity=1, line width=1, stealth-stealth] (-5,0) -- (-5,100) ;
	\draw [draw opacity=1, line width=1, stealth-stealth] (0,-5) -- (200,-5) ;
	
	\draw[draw opacity=1, line width=1, -stealth]   (100,60) -- (100,80) ;
	\draw[draw opacity=1, line width=1, -stealth]   (100,60) -- (120,60) ;

	
	\draw (100,-7) node [anchor=north west][inner sep=0.75pt]    {$L$};
	\draw (-16,50) node [anchor=north west][inner sep=0.75pt]    {$H$};
	\draw (105,58) node [anchor=north west][inner sep=0.75pt]    {$x$};
	\draw (93,70) node [anchor=north west][inner sep=0.75pt]    {$y$};
	
	\draw (10,75) node [anchor=north west][inner sep=0.75pt]    {Fluid $h$};
	\draw (10,30) node [anchor=north west][inner sep=0.75pt]    {Fluid $l$};
	
\end{tikzpicture}
	\caption{Schematic of the test domain used to simulate the thermocapillary-driven motion of two-fluids within a heated microchannel.}
	\label{fig:layeredFlow_schematic}
\end{figure}

The test domain for this problem is depicted in \cref{fig:layeredFlow_schematic}. Here, it can be seen that the channel has a length of $L$, a height of $H$, and is split horizontally into two even sections of $\phi_L=0$ and $\phi_H=1$ corresponding to a system with two fluids with their own respective densities. Wall boundary conditions are imposed at the bottom and top of the domain with Dirichlet temperature conditions defined as,

\begin{align}
	\label{eq::dirichlet_temp_b}
	T(x,-H/2) &= T_h + T_0\cos{\omega x}, \\
	\label{eq::dirichlet_temp_t}
	T(x,H/2)  &= T_c,
\end{align}

\noindent where the temperatures are specified such that $0<T_0<T_c<T_h$, and $\omega=2\phi/L$ represents the wave number.

To define the parameters of the solution, the relevant non-dimensional numbers are introduced,

\begin{align}
	\Ma &= \frac{\rho_H c_{p,H}LU}{k_H}, \\
	\Rey &= \frac{\rho_H UL}{\mu_H}, \\
	\Ca &= \frac{U\mu_H}{\sigma_{ref}}, 
\end{align}

as well as the ratios of density $\rho^*$, viscosity $\mu^*$, heat conduction $k^*$, and heat capacity $c_p^*$. For this problem, the characteristic length and velocity are equal to $L$, and $U=b\left|\sigma_T\right|/\mu_H L$, respectively. To further simplify, parameters are chosen such that $\Rey$, $\Ca$, and $\Ma$ are sufficiently small which means that the convective transport of momentum, and energy can be neglected. As such, the interface in the simulation remains planar, and is initialised as,
\begin{equation}
	\phi = \phi_0 + \frac{\phi_H - \phi_L}{2}\tanh\left(\frac{y}{\xi /2}\right),\quad -b<y<a.
\end{equation}
Here, the values of $a$ and $b$ represent the height of the top and bottom fluid layers, and are taken as $a=b=H/2$. The analytical solution for this flow is provided in Appendix~A.

For the numerical study, the simulations were specified with a characteristic length of $L = 256$ lattice cells. This results in a channel of $512 \times 256$ cells and $a = b = 128$.  Periodic boundary conditions are applied to the left and right of the channel for all fields. For the phase-field and the hydrodynamic \ac{PDF}s, no-slip boundary conditions are applied on the upper and lower walls by applying the bounce-back method. For the thermal \ac{PDF}s, Dirichlet boundary conditions are applied according to \cref{eq::dirichlet_temp_b,eq::dirichlet_temp_t}. The wall temperatures are specified as $T_h = 20$, $T_c = 10$ and $T_0 = 4$, respectively. The numerical properties for the two fluids are chosen as $\sigma_T = -5 \cdot 10^{-4}$, $\sigma_{\mathrm{ref}} = 2.5 \cdot 10^{-2}$, $\nu_{\mathrm{A, B}} = 0.2$, $\kappa_{\mathrm{B}} = 0.2$ and  $M_{\phi} = 5 \cdot 10^{-2}$. Two test cases are constructed by setting $\kappa_{\mathrm{A}} = 0.2$ and $0.04$, corresponding to heat conduction ratios of $k^*=1$ and $k^* = \nicefrac{1}{5}$, respectively.

\begin{figure}[htb]
	\centering
	\input{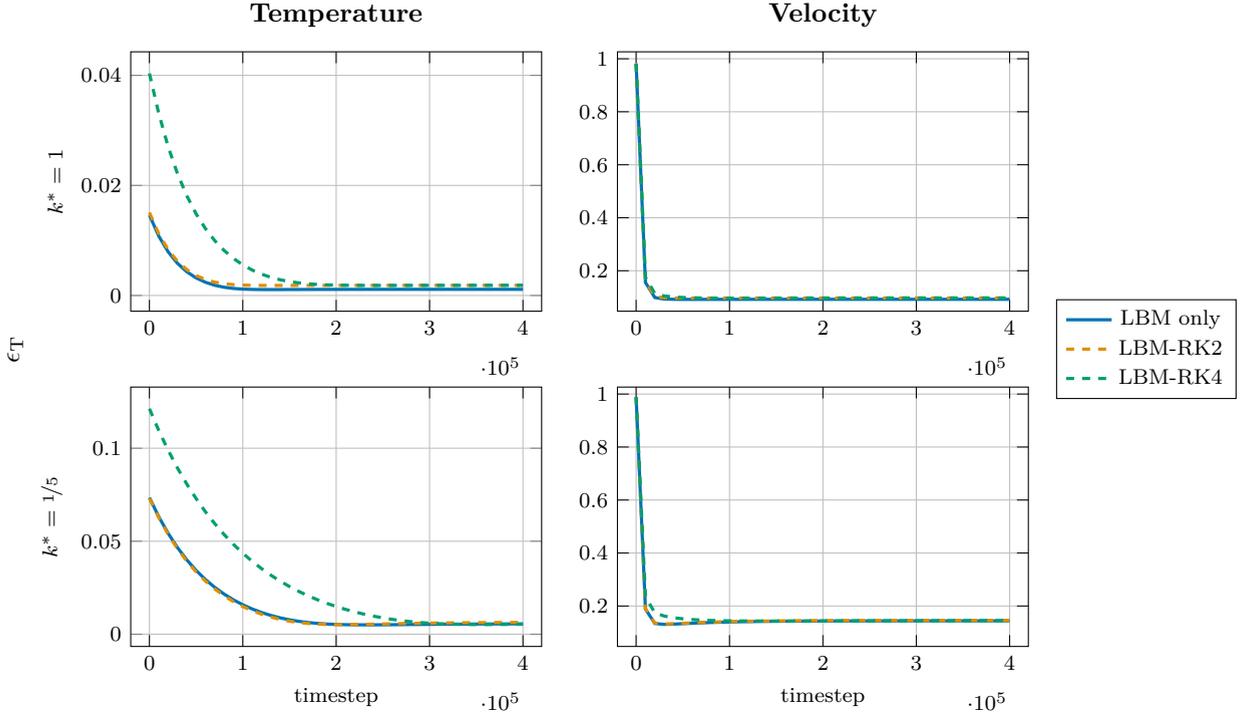}	
	\caption{Convergence of the temperature and velocity fields using a \ac{LBM} scheme for all order parameters (\ac{LBM}-only)  with comparison to using \ac{RK2} and \ac{RK4} scheme to obtain the temperature respectively.}
	\label{fig:conv_2d}
\end{figure}

To examine the error present in each solution, an $\ell_2$-norm is defined as,
\begin{equation}
	\ell_2 = \sqrt{\frac{\sum_{i,j} (|\mathbf{\psi}_{i,j}^{num}| - |\mathbf{\psi}_{i,j}^{A}|)^2}{\sum_{i,j}|\mathbf{\psi}_{i,j}^{A}|^2}},
\end{equation}
where $\psi$ is a placeholder for temperature or velocity, and the superscript indicates if the value was obtained numerically or from the analytical solution given in \ref{sec:anlytical-solution}. The simulation is run for $4 \cdot 10^5$ timesteps and the evolution of the errors for the temperature and the velocity are shown in \cref{fig:conv_2d}.  

As expected, switching the numerical scheme to obtain the temperature does not significantly impact the evolution of the velocity field. However, the behaviour of the temperature field shows variation in behavior, especially in the second configuration (i.e. $k^*=\nicefrac{1}{5}$). Furthermore, while all schemes converge almost to the same accuracy, it is noticeable that the \ac{RK4} scheme had slower transient behaviour, taking more iterations to convergence to the steady-state result. The magnitude of errors at steady state are provided in \cref{tab:planarflowDP2D}, where the difference between selected methods is only minor.

\begin{table}[h!]
	\centering
	\caption{Comparison of errors for the temperature and the velocity field after $4 \cdot 10^5$ timestep}
	\label{tab:planarflowDP2D}
	\begin{tabular}{ l c c c c}
		\hline
		\textit{$\ell_2$-norm}  & \multicolumn{2}{c}{$k^*= 1$}         & \multicolumn{2}{c}{$k^*=\nicefrac{1}{5}$}        \\
		\textbf{Model setup} & \textit{Temperature} & \textit{Velocity} & \textit{Temperature}  & \textit{Velocity} \\
		\hline
		LBM only & 1.1164e-3 & 9.2165e-2 & 5.5609e-3 & 1.4398e-1 \\ 
		LBM-RK2 & 1.8479e-3 & 9.7223e-2 & 6.3654e-3 & 1.4548e-1 \\ 
		LBM-RK4 & 1.8457e-3 & 9.7219e-2 & 5.3611e-3 & 1.4518e-1 \\ 
		\hline
	\end{tabular}
\end{table}

\begin{figure}[htb]
	\centering
	\subfloat[$k^*= 1$]{
	   \includegraphics[trim={60, 20, 70, 40},clip,width=.45\linewidth]   {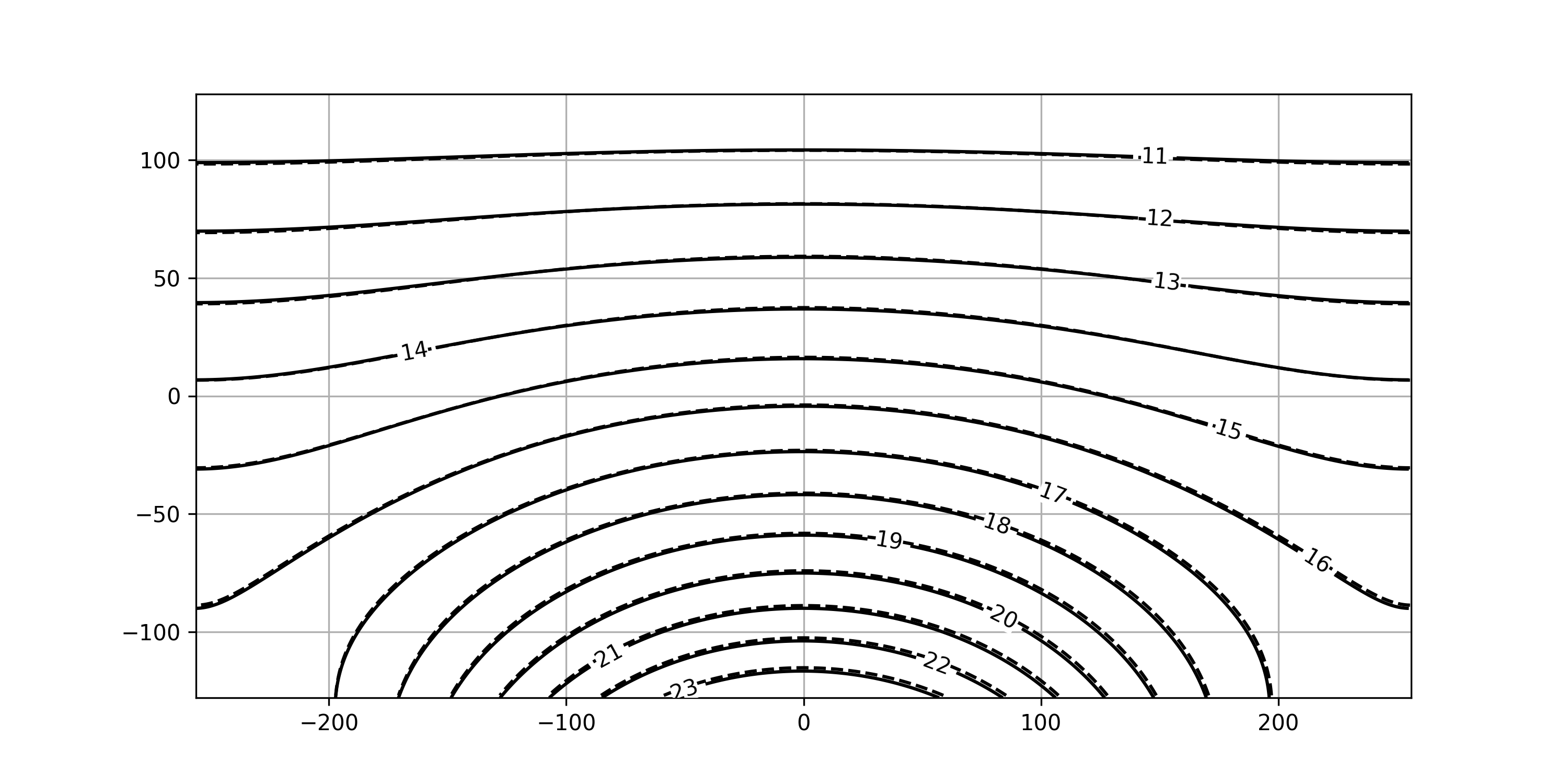}
	\label{fig:temperature_contour_cm_cm_cm_dp_case_1} 
    }
	\subfloat[$k^*=\nicefrac{1}{5}$]{
	   \includegraphics[trim={60, 20, 70, 40},clip,width=.45\linewidth]   {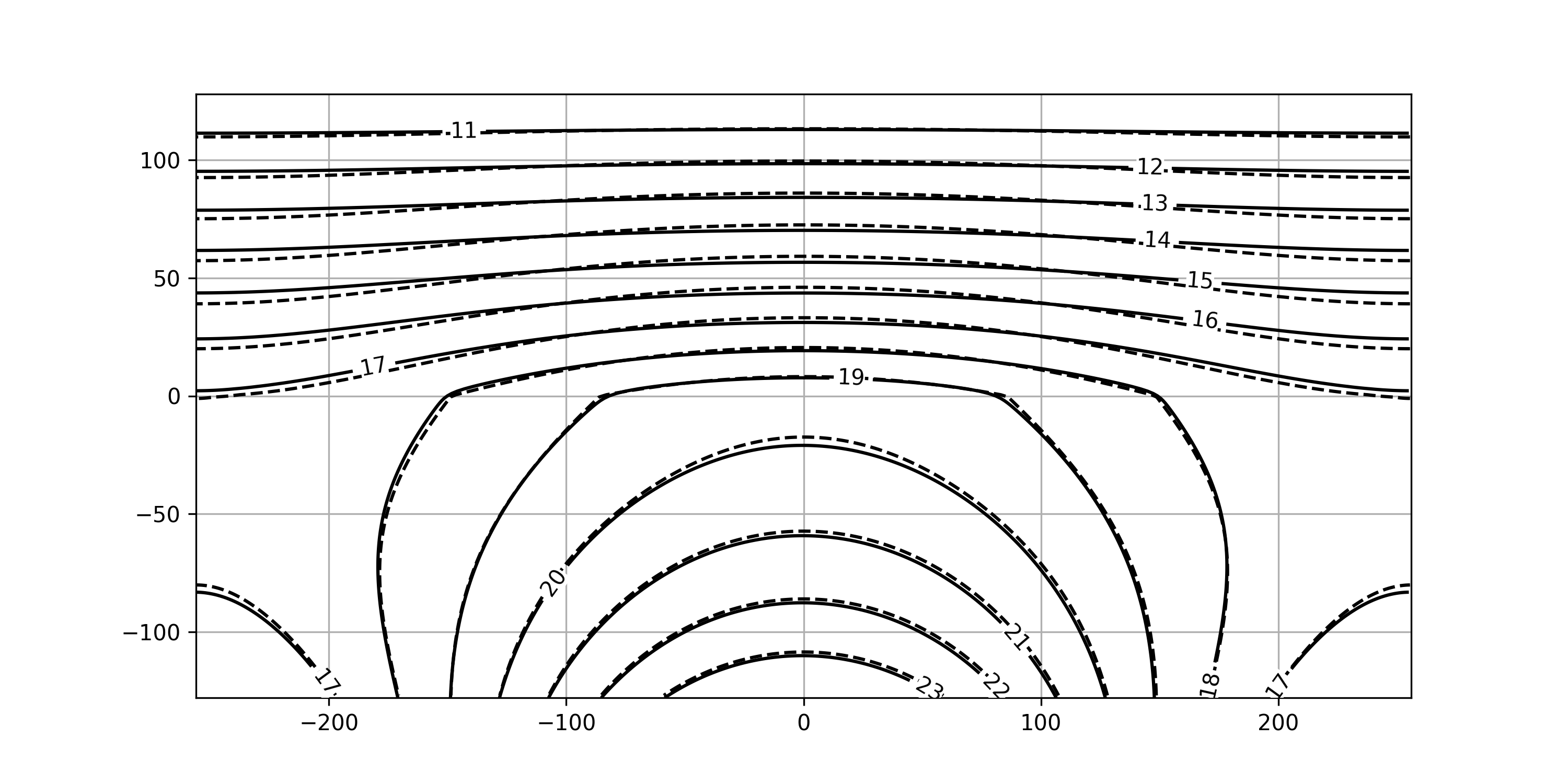}
	\label{fig:temperature_contour_cm_cm_cm_dp_case_2} 
    }
	\caption{Temperature contours of fluid systems with heat conduction ratio (a) $k^* = 1$ and (b) $k^* = \nicefrac{1}{5}$. Analytical solutions are represented using solid lines while the simulation results are shown with dashed lines. The results were conducted with a pure LBM-based method. Similar results could be achieved with a hybrid method, where the energy equation was solved using an RK2 or an RK4 scheme.}
	\label{fig:temperature_contour_2D}
\end{figure}

A contour plot of the temperature field against the analytical solution is provided in \cref{fig:temperature_contour_2D}. Due to the insignificant difference between the three variants, only the pure LBM-based solution is shown.

\subsubsection{Pseudo-3D analysis for impact of lattice structure}

To test the methodology in a three-dimensional setup, the planar heated channel was extended to a pseudo-3D form. For the pseudo-3D setup the same parameters as in \cref{sec:planarChanel2D} were used. However, the domain was extended by ten cells in the $z$-direction. This procedure does not yet cover real complex three-dimensional flow domains. However, it forms a simple test bed where the same analytical solutions can be used for a cut plane in the $z$-direction. This is advantageous because analytical solutions for complex flow scenarios such as thermocapillary flows can prove challenging. 

The pseudo-3D simulations were carried out using a D3Q27 stencil with a central-moment \ac{LBM} scheme for the hydrodynamic and Allen-Cahn equations. For the energy equation, a central-moment \ac{LBM} was employed with various lattice stencils, namely, D3Q7, D3Q15, D3Q19 and D3Q27. These lattice stencils are shown in greater detail in \ref{sec:stencils}. While reduced stencils are established in literature for solving diffusion equations with the \ac{LBM}, to the best of the authors' knowledge, it has not yet been demonstrated for thermocapillary flows and is thus analysed here. Furthermore, it is important to note that reduced stencils, while showing accurate results, can impact the stability of the solver negatively for complex flow dynamics \cite{mitchell2021computational}.

\begin{figure}[htb]
	\centering
	\input{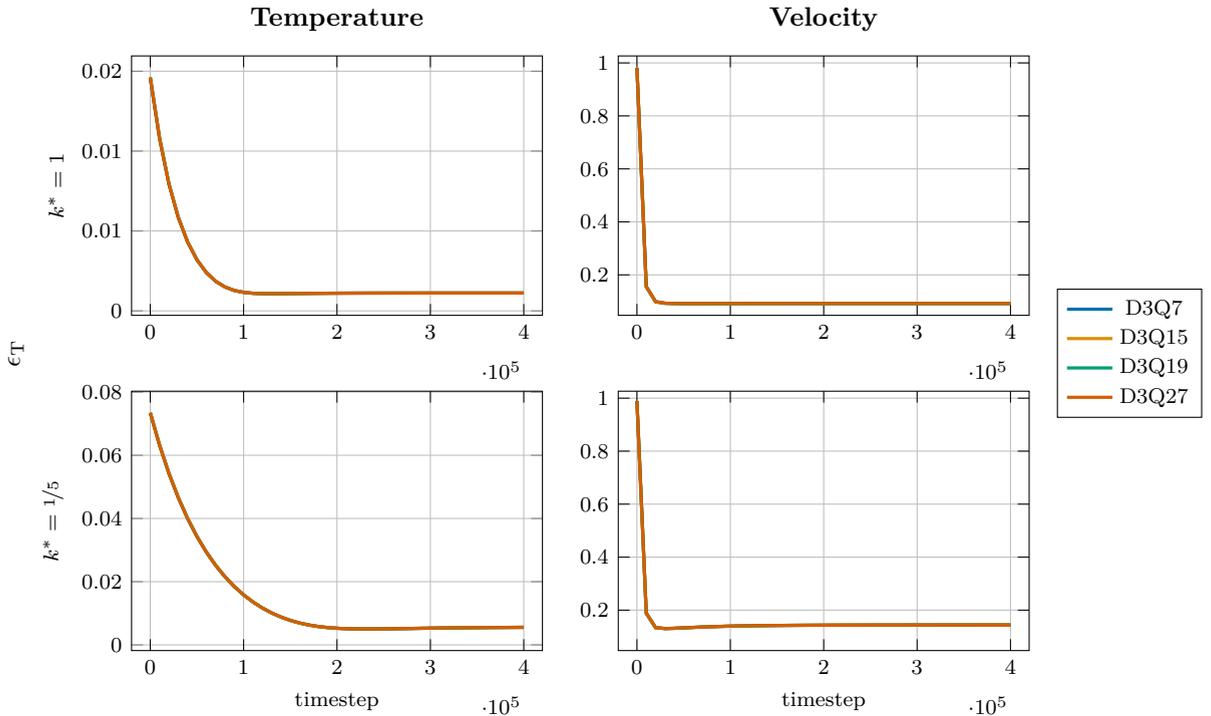}	
	
	\caption{Convergence of the temperature and velocity field for different stencils for the thermal PDFs for the pseudo-3D simulation. The results of all stencils overlap, without any noticeable difference.}
	\label{fig:conv_3d}
\end{figure}

Similarly to \cref{sec:planarChanel2D}, the simulation results were compared to the analytical solution from \ref{sec:anlytical-solution}. Due to the three-dimensional setup, a two-dimensional cross-section at $z = 5$ lattice cells was used. The evolution of the error for the temperature as well as for the velocity are plotted in \cref{fig:conv_3d} and the final errors after $4 \cdot 10^5$ timesteps are listed in \cref{tab:planarflowDP3D}. It is noticeable that no difference between the stencils was obtained. While this does not yet show the feasibility for complex real 3D flow dynamics, it can already function as a guideline. Thus, using the D3Q7 stencil seems to be promising for simple thermocapillary flows, allowing minimal computational cost without significant impact on simulation results.

\begin{table}[htb]
	\centering
	\caption{Comparison of errors for the temperature and the velocity field in the pseudo-3D simulations after $4 \cdot 10^5$ timestep with varying lattice stencils for the \ac{LBM} heat solver.}
	\label{tab:planarflowDP3D}
	\begin{tabular}{ l c c c c}
		\hline
		\textit{$\ell_2$-norm}  & \multicolumn{2}{c}{$k^*= 1$} &    \multicolumn{2}{c}{$k^*=\nicefrac{1}{5}$}     \\
		\textbf{Lattice Stencil} & \textit{Temperature} & \textit{Velocity} & \textit{Temperature}  & \textit{Velocity} \\
		\hline
		D3Q7 & 1.1165e-3 & 9.2094e-2 & 5.5628e-3 & 1.4389e-1 \\ 
		D3Q15 & 1.1164e-3 & 9.2047e-2 & 5.5607e-3 & 1.4385e-1 \\ 
		D3Q19 & 1.1164e-3 & 9.2047e-2 & 5.5606e-3 & 1.4385e-1 \\ 
		D3Q27 & 1.1164e-3 & 9.2047e-2 & 5.5607e-3 & 1.4385e-1 \\ 
		
		\hline
	\end{tabular}
\end{table}

\subsection{Droplet motion with local laser heating in 2D}
\label{sec:droplet2D}

With the impact of solution methodology analysed in the previous sections, here, the model is now applied to provide insight on capturing droplet dynamics in a laser heated channel setup. This is done in two steps. First, the developed methodology is applied to a two-dimensional droplet for comparison with existing literature, in particular the studies of Liu~et~al.~\cite{liu_valocchi_zhang_kang_2014}. After obtaining confidence in solving thermocapillary droplet dynamics, the setup was extended to three-dimensional flows to study the effect in a more realistic scenario. To replicate the thermal effects induced by a laser, a heat source was introduced within the channel as,

\vspace{-1cm}
\begin{align}
	\label{eq:HeatSource}
	q_T =
	\begin{cases}
		Q_s \exp{\left(-2 \frac{\left(x - x_s\right)^2 + \left(y - y_s\right)^2 + \left(z - z_s\right)^2}{w_s^2}\right)}, & \text{if } \left[\left(x - x_s\right)^2 + \left(y - y_s\right)^2 + \left(z - z_s\right)^2\right] \geq d_s^2 \\
		0, & \text{otherwise}
	\end{cases}.
\end{align}

\noindent Here, $Q_s$ signifies the maximum heat flux generated by the laser, while $x_s$, $y_s$, and $z_s$ denote the precise laser position.  For the two-dimensional cases, the $z$-component was disregarded. The extent of heat dispersion is defined by $d_s$, while $w_s$ serves as a key parameter governing the heat flux profile. This particular experimental configuration mirrors the methodology employed by Liu~et~al.~\cite{liu_valocchi_zhang_kang_2014}.

Within this framework, a semicircular droplet with a radius of $R = 32$ is situated at the coordinates $x_c, y_c = (65, 0)$ within a computational domain of dimensions $L_x \times L_y = 8R \times 2R$. Initially, the droplet is stationary, with no initial velocity field defined. To induce motion in the channel, a velocity of $U_w$ is applied to the upper wall, while the lower wall remains stationary, resulting in a constant shear rate of $\gamma = \frac{U_w}{L_y}$.

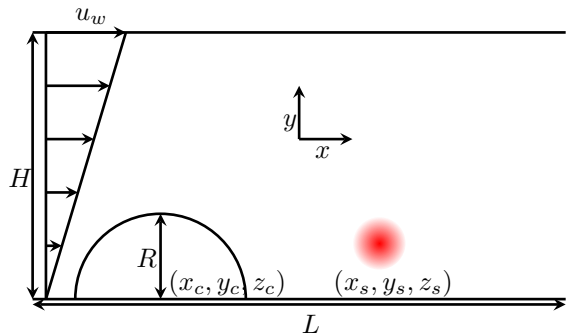
\begin{figure}[htb]
	\centering
	\begin{tikzpicture}[x=1pt, y=1pt]
	
	\draw [draw opacity=1 ][line width=1]    (0,0) -- (200,0) ;
	\draw [draw opacity=1 ][line width=1]    (0,100) -- (200,100) ;
	
	\draw [draw opacity=1, line width=1, stealth-stealth] (0,0) -- (0,100) ;
	\draw [draw opacity=1, line width=1, stealth-stealth] (0,-2) -- (200,-2) ;
	
	\draw[draw opacity=1, line width=1, -stealth]   (100,60) -- (100,80) ;
	\draw[draw opacity=1, line width=1, -stealth]   (100,60) -- (120,60) ;
	
	\draw  [draw opacity=1 ][line width=1]  (5,0) -- (5,100) ;
	\draw  [draw opacity=1 ][line width=1]  (5,0) -- (35,100) ;
	
	\draw  [draw opacity=1, line width=1, -stealth]  (5,20) -- (11,20) ;
	\draw  [draw opacity=1, line width=1, -stealth]  (5,40) -- (17,40) ;
	\draw  [draw opacity=1, line width=1, -stealth]  (5,60) -- (23,60) ;
	\draw  [draw opacity=1, line width=1, -stealth]  (5,80) -- (29,80) ;
	\draw  [draw opacity=1, line width=1, -stealth]  (5,100) -- (35,100) ;

	\draw (15,110) node [anchor=north west][inner sep=0.75pt]    {$u_w$};
	
	\draw (100,-5) node [anchor=north west][inner sep=0.75pt]    {$L$};
	\draw (-10,50) node [anchor=north west][inner sep=0.75pt]    {$H$};
	\draw (105,58) node [anchor=north west][inner sep=0.75pt]    {$x$};
	\draw (93,70) node [anchor=north west][inner sep=0.75pt]    {$y$};
	
	\draw[thick,black,line width=1] (80,0) arc (0:180:32);
	\draw  [draw opacity=1, line width=1, stealth-stealth]  (48,0) -- (48,32) ;
	
	\tikzset
	{
		myCircle/.style=
		{
			red,
			path fading=fade out,
		}
	}
	
	\fill[myCircle] (130,21) circle (10);
	
	
	\draw (38,20) node [anchor=north west][inner sep=0.75pt]    {$R$};
	\draw (50,12) node [anchor=north west][inner sep=0.75pt]    {$(x_c, y_c, z_c)$};
	
	\draw (112,12) node [anchor=north west][inner sep=0.75pt]    {$(x_s, y_s, z_s)$};

\end{tikzpicture}
	\caption{Schematic of the test domain used to simulate the thermocapillary-driven motion of a droplet in a channel.}
	\label{fig:droplet_setup_schematic}
\end{figure}

In addition, temperature boundary conditions are set to zero ($T_{\mathrm{ref}} = 0$) for both the bottom and top walls, while periodic boundary conditions are applied to the west and east walls. Notably, the temperature of the fluid is solely influenced by a laser located at $x_s = 181$ and $y_s = 21$, and characterised by parameters $w_s = 6.0$, $d_s = 8.0$, and $Q_s = 0.2$. A schematic representation of the experimental setup is presented in \cref{fig:droplet_setup_schematic}.

The remaining flow parameters are established by their respective dimensionless counterparts, which are explicitly defined as $\Ma = 0.01$, $\Ca = 0.16$, and $\Rey = 0.08$. The dimensionless conductivity ratio of the fluids is represented as $k^* = 1$. Additionally, the surface tension is characterised by the values of $\sigma_T = 2 \times 10^{-4}$ and $\sigma_{ref} = 5 \times 10^{-3}$. The velocity of the upper wall, denoted as $U_w$, can be determined based on the characteristic lengths of the system, where $L = R$ and $U = \gamma R$. To specify the interface thickness, $\xi$, a discretisation of four cells was employed. Similarly, the dimensionless time was defined as $t^* = \gamma t$
  
Variations in the wetting angle of the droplet were investigated, and its trajectory was tracked to evaluate the impact of the localised heating source. The implementation of this boundary condition was applied based on that proposed by Fakhari~et~al.~\cite{FAKHARI2017620}, for which the performance and efficient 3D implementation options are discussed in Sashko~et~al.~\cite{sashko2023}.

\begin{figure}[htb]
    \centering
    \input{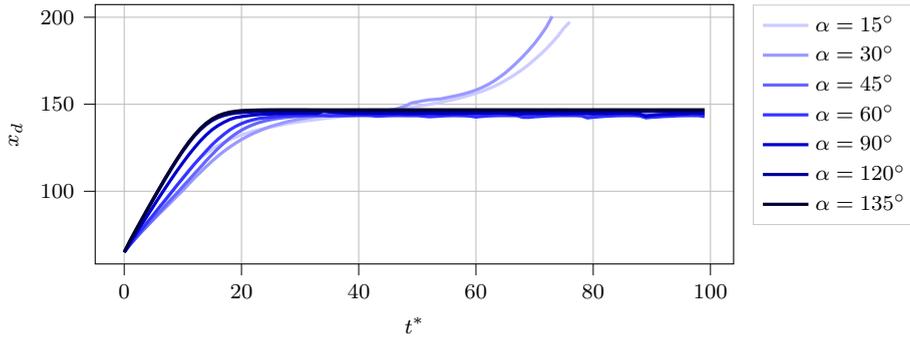}
    \caption{Droplet migration in 2D for different contact angles. The plot shows the $x$-coordinate of the droplet centroid over the dimensionless time $t^*$. Two droplets with angles 15$^\circ$ and 30$^\circ$ facilitate droplet motion through the local heat source. Droplets with higher contact angle to the wall are completely blocked by the introduced heat source.}
    \label{fig:dropletmigration2d}
\end{figure}

This investigation closely parallels the approach adopted in the study by Liu et al.~\cite{liu_valocchi_zhang_kang_2014}. The observed behaviours of the droplets are visually presented in \cref{fig:dropletmigration2d}, where most droplets are blocked by the laser-heated focal point. However, for contact angles measuring 15$^\circ$ and 30$^\circ$, the augmented wetting properties of the droplet proved sufficiently dominant to overcome the asymmetrical forces acting across the interface, thereby facilitating the continued motion of the droplet. It is notable that in the study by Liu et al.~\cite{liu_valocchi_zhang_kang_2014}, droplets with a contact angle of 45$^\circ$ were not blocked, while the same droplets are blocked in the results presented here. There could be various reasons for this, including the fact that the prior study used a Cahn-Hilliard \cite{Hilliard58} approach to track the interface between the fluids, while here the conservative Allen-Cahn equation is employed.

With this planar representation of the droplet behaviour, one is neglecting any potential for the droplet to move in the transverse direction and propagate along the circumference of the heat source. To investigate this behaviour, the 2D domain was extended to a three-dimensional configuration.

\subsection{Three-dimensional impacts on droplet motion with local laser heating}\label{sec:3d-laser}

Extending the previous analysis, as well as those found in the literature, this section analyses the behaviour of a three-dimensional thermocapillary droplet in a shear flow. This facilitates the degrees of freedom necessary to see motion perpendicular to the primary shear direction, and increased complexity in the droplet-heated spot interaction. For this analysis, the same dimensionless numbers as in \cref{sec:droplet2D} were used. In order to avoid any boundary effects, the domain size was specified as $L_x \times L_y \times L_z = 16R \times 2R \times 8R$. A droplet was placed in the domain at rest at $x_c, y_c, z_c = (2R+1, 0, 4R)$ and the radius of the droplet was prescribed as $R = 32$ lattice cells. 

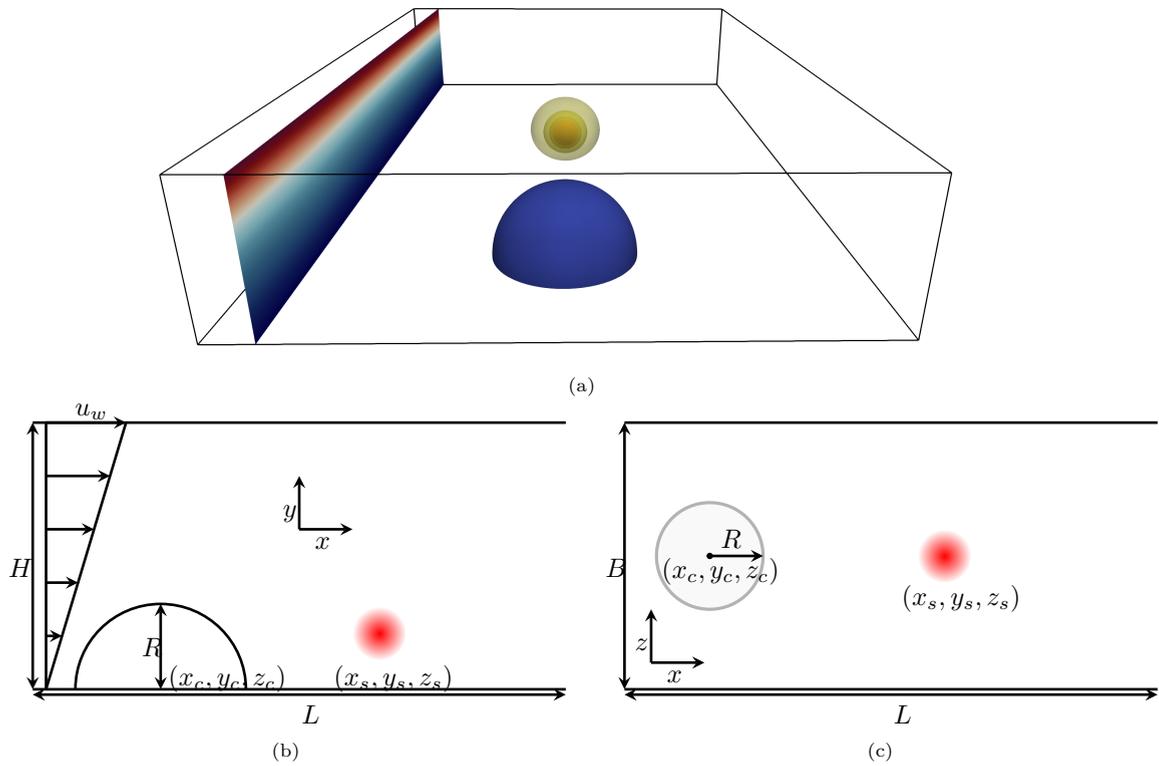
\begin{figure}
	\centering
	\subfloat[]{
	\includegraphics[trim={150 0 150 0},clip, width=0.7\linewidth]{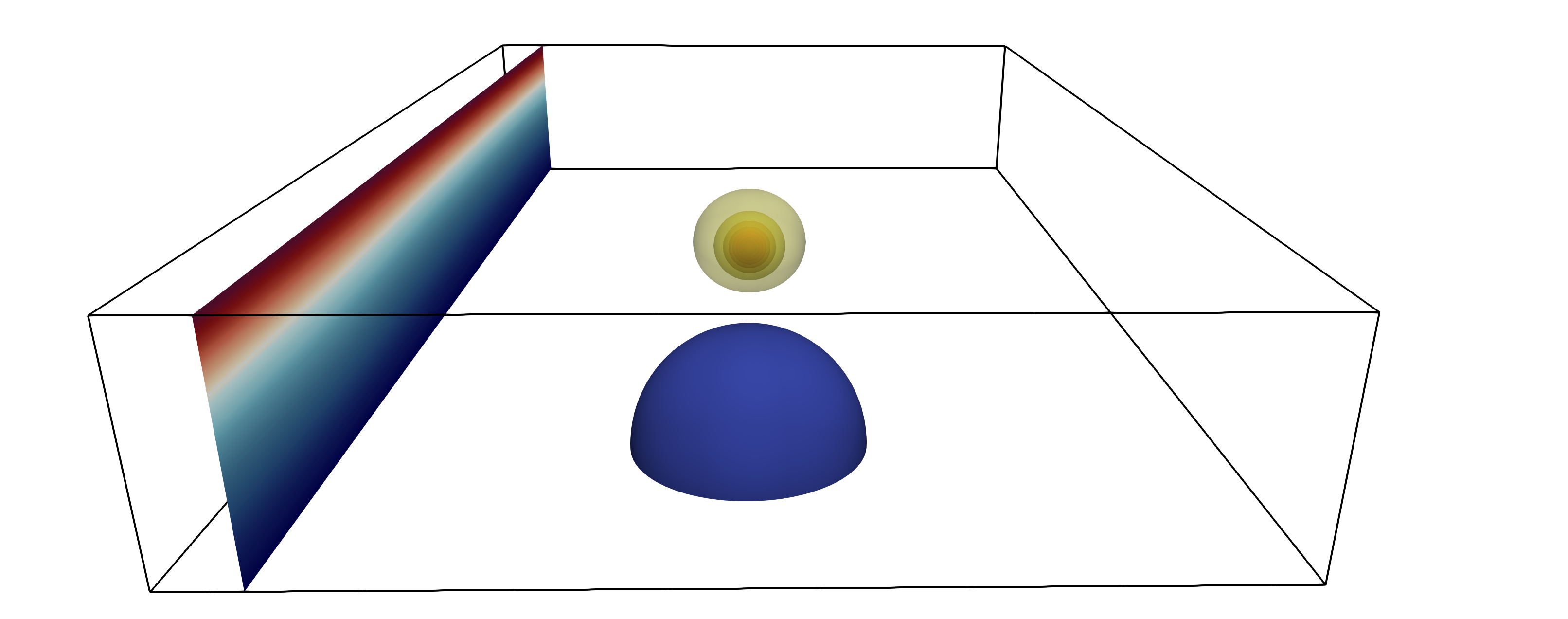}
	\label{fig:dropletSetupParaview}
	}

	\subfloat[]{
	    \begin{tikzpicture}[x=1pt, y=1pt]
	
	\draw [draw opacity=1 ][line width=1]    (0,0) -- (200,0) ;
	\draw [draw opacity=1 ][line width=1]    (0,100) -- (200,100) ;
	
	\draw [draw opacity=1, line width=1, stealth-stealth] (0,0) -- (0,100) ;
	\draw [draw opacity=1, line width=1, stealth-stealth] (0,-2) -- (200,-2) ;
	
	\draw[draw opacity=1, line width=1, -stealth]   (100,60) -- (100,80) ;
	\draw[draw opacity=1, line width=1, -stealth]   (100,60) -- (120,60) ;
	
	\draw  [draw opacity=1 ][line width=1]  (5,0) -- (5,100) ;
	\draw  [draw opacity=1 ][line width=1]  (5,0) -- (35,100) ;
	
	\draw  [draw opacity=1, line width=1, -stealth]  (5,20) -- (11,20) ;
	\draw  [draw opacity=1, line width=1, -stealth]  (5,40) -- (17,40) ;
	\draw  [draw opacity=1, line width=1, -stealth]  (5,60) -- (23,60) ;
	\draw  [draw opacity=1, line width=1, -stealth]  (5,80) -- (29,80) ;
	\draw  [draw opacity=1, line width=1, -stealth]  (5,100) -- (35,100) ;

	\draw (15,108) node [anchor=north west][inner sep=0.75pt]    {$u_w$};
	
	\draw (100,-5) node [anchor=north west][inner sep=0.75pt]    {$L$};
	\draw (-10,50) node [anchor=north west][inner sep=0.75pt]    {$H$};
	\draw (105,58) node [anchor=north west][inner sep=0.75pt]    {$x$};
	\draw (93,70) node [anchor=north west][inner sep=0.75pt]    {$y$};
	
	\draw[thick,black,line width=1] (80,0) arc (0:180:32);
	\draw  [draw opacity=1, line width=1, stealth-stealth]  (48,0) -- (48,32) ;
	
	\tikzset
	{
		myCircle/.style=
		{
			red,
			path fading=fade out,
		}
	}
	
	\fill[myCircle] (130,21) circle (10);
	
	
	\draw (40,20) node [anchor=north west][inner sep=0.75pt]    {$R$};
	\draw (50,10) node [anchor=north west][inner sep=0.75pt]    {$(x_c, y_c, z_c)$};
	
	\draw (112,10) node [anchor=north west][inner sep=0.75pt]    {$(x_s, y_s, z_s)$};

\end{tikzpicture}
		\label{fig:dropletSetupSchemaXY} 
	}
	\subfloat[]{
		\begin{tikzpicture}[x=1pt,y=1pt]
	
	\draw [draw opacity=1 ][line width=1]    (0,0) -- (200,0) ;
	\draw [draw opacity=1 ][line width=1]    (0,100) -- (200,100) ;
	
	\draw [draw opacity=1, line width=1, stealth-stealth] (0,-2) -- (200,-2) ;
	\draw[draw opacity=1, line width=1, stealth-stealth]   (0,0) -- (0,100) ;
	
	\draw[draw opacity=1, line width=1, -stealth]   (10,10) -- (10,30) ;
	\draw[draw opacity=1, line width=1, -stealth]   (10,10) -- (30,10) ;

	\filldraw[color=gray!60, fill=gray!5, very thick](32,50) circle (20);
	
	\draw[draw opacity=1, line width=1, -stealth]   (32,50) -- (52,50) ;
	\draw[draw opacity=1, fill=black]   (32,50) circle (1) ;
	
	\tikzset
	{
		myCircle/.style=
		{
			red,
			path fading=fade out,
		}
	}
	
	\fill[myCircle] (120,50) circle (10);

	
	\draw (100,-5) node [anchor=north west][inner sep=0.75pt]    {$L$};
	\draw (-8,50) node [anchor=north west][inner sep=0.75pt]    {$B$};
	\draw (15,8) node [anchor=north west][inner sep=0.75pt]    {$x$};
	\draw (3,20) node [anchor=north west][inner sep=0.75pt]    {$z$};
	
	\draw (40,52) node [above][inner sep=0.75pt]    {$R$};
	\draw (13,50) node [anchor=north west][inner sep=0.75pt]    {$(x_c, y_c, z_c)$};
	\draw (103, 40) node [anchor=north west][inner sep=0.75pt]    {$(x_s, y_s, z_s)$};

\end{tikzpicture}
		\label{fig:dropletSetupSchemaXZ} 
	}

\caption{Configuration of the test domain for the three-dimensional extension of the simulation of a thermocapillary droplet, showing (a) the initial setup of the droplet in a shear flow channel, (b) an elevation of the domain in the $xy$-plane, and (c) a plan view of the domain in the $xz$-plane. Similar to the analysis in \cref{sec:droplet2D}, a single heat source is introduced to the thermal solver.}
	
\end{figure}

A three-dimensional view of the initial setup is shown in \cref{fig:dropletSetupParaview}. Here, the temperature field is already evolved for $t^*$ timesteps which is the number of timesteps used for the temperature initialisation. At this point, the velocity field, as well as the phase-field are still at rest. To better illustrate the droplet behaviour, both $xy$- and a $xz$-plane schematics are presented in \cref{fig:dropletSetupSchemaXZ} and \cref{fig:dropletSetupSchemaXY}.

\begin{figure}[htb]
	\centering
	\subfloat[]{
		\includegraphics[trim={150 0 150 0},clip, width=0.7\linewidth]{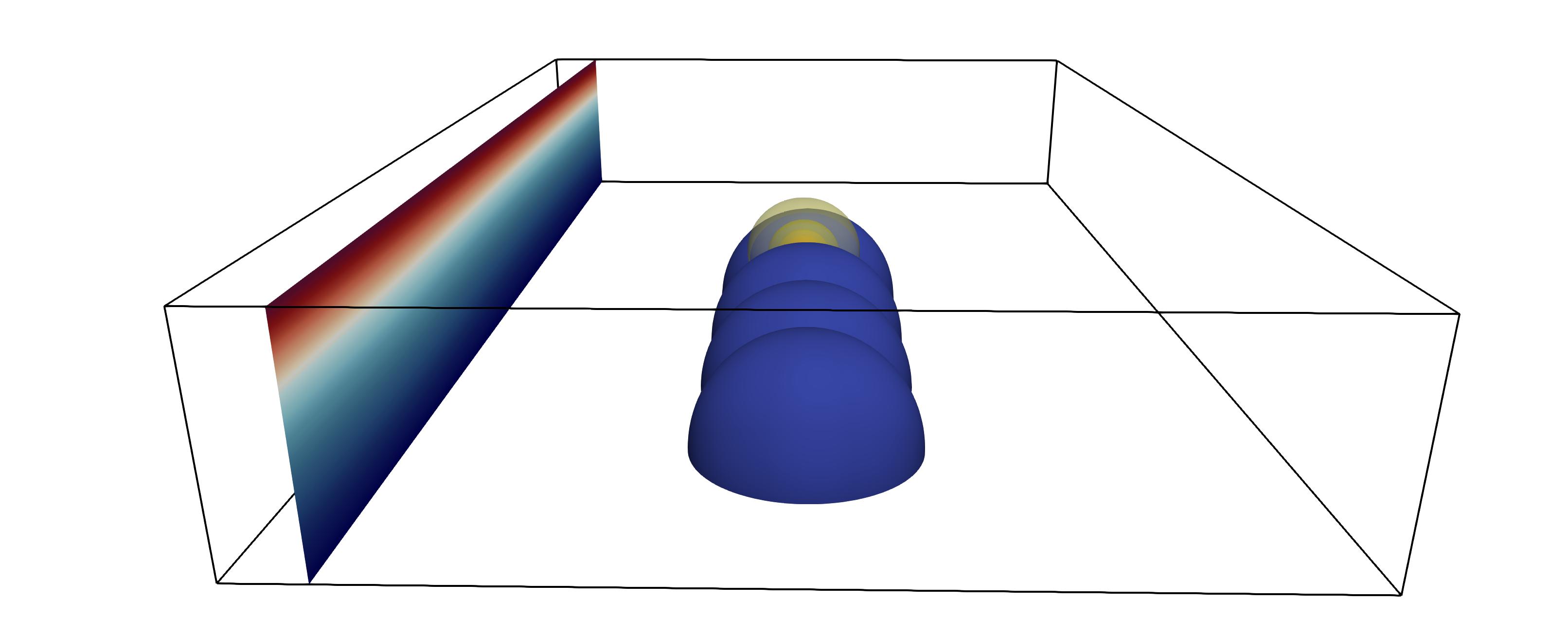}
		\label{fig:singleHeatSourceCase1Paraview}
	}
	
	\subfloat[]{
		\label{fig:singleHeatSourceCase1Plot}
		\input{figures/droplet_migration/3D/singleHeatSource_heat_flux_20.tex}
	}

\caption{Droplet migration in a three-dimensional channel flow with various contact angles, showing (a) the evolution of a droplet with $90^\circ$ contact angle where each droplet in the figure represents the movement during a period of $t^* = 10$ and (b) the $x$- and $z$-coordinate of the centroid of droplets with various contact angles as a function of the dimensionless time respectively. For the laser point, a maximal heat flux of $Q_s = 0.2$ was used.}
\label{fig:singleHeatSourceCase1}
\end{figure}

The simulation was carried out with different contact angles, similarly to the two-dimensional test case. These contact angles were varied from $45^\circ$ to $135^\circ$. The results of the numerical simulation are shown in \cref{fig:singleHeatSourceCase1}. Therein, \cref{fig:singleHeatSourceCase1Plot} shows the displacement in the $x$- and $z$-plane of various droplets with different contact angles, while \cref{fig:singleHeatSourceCase1Paraview} presents an example of the evolution of a droplet with a $90^\circ$ contact angle. Unlike the two-dimensional simulation, all droplets can pass the laser point and none of them can be stopped. Here, the impact of a spherical heat source on the semi-spherical droplet in a three-dimensional domain is not replicated by the two-dimensional planar analogue.

\begin{figure}[htb]
	
	\centering
	\subfloat[]{
		\includegraphics[trim={150 0 150 0},clip, width=0.7\linewidth]{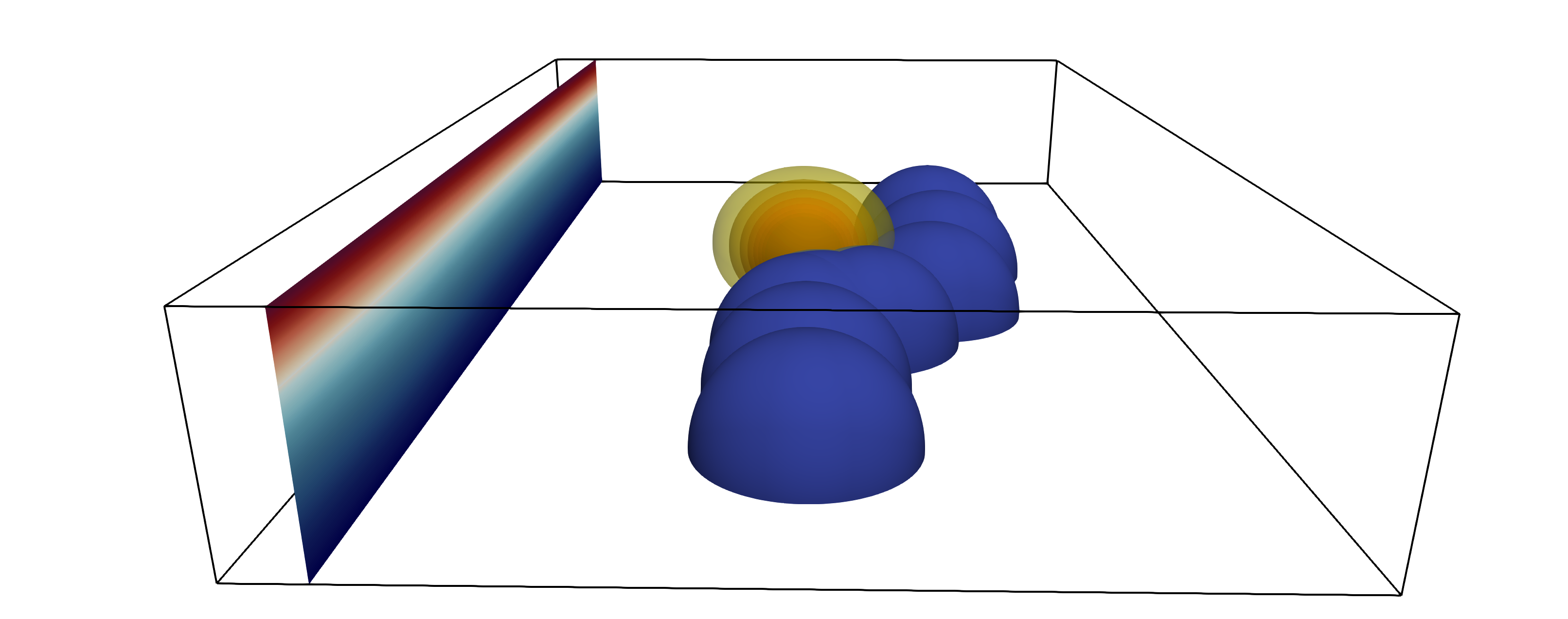}
		\label{fig:singleHeatSourceCase2Paraview}
	}
	
	\subfloat[]{
		\label{fig:singleHeatSourceCase2Plot}
		\input{figures/droplet_migration/3D/singleHeatSource_heat_flux_100.tex}
	}
\caption{Droplet migration in a three-dimensional channel flow with various contact angles, showing (a) evolution of a droplet with $90^\circ$ contact angle where each droplet in the figure represents the movement during a period of $t^* = 10$ and (b) the $x$- and $z$-coordinate of the centroid of droplets with various contact angles. For the laser point, a maximal heat flux of $Q_s = 1$ was used.}
\label{fig:singleHeatSourceCase2}
\end{figure}

As a next step, the test case was modified by increasing the heat flux by a factor of five to $Q_s = 1$. This was done to investigate if capture can be achieved with a higher intensity laser, which would generate additional heat flux. The results of the second simulation are shown in \cref{fig:singleHeatSourceCase2}. Therein, \cref{fig:singleHeatSourceCase2Plot} shows the displacement in the $x$- and $z$-plane of various droplets with different contact angles, while \cref{fig:singleHeatSourceCase2Paraview} presents an example of the evolution of a droplet with a $90^\circ$ contact angle. It can be observed that the droplets move towards the laser point until $\approx t^* = 10$. At this time, all droplets seem to stop in a manner similar to the two-dimensional case. However, after a period of time (depending on the wetting angle), all droplets start to increase their velocity and pass the heated laser point in the $z$-plane. This is pictured clearly in \cref{fig:singleHeatSourceCase2Plot}, where the $z$-coordinate of the centroid of the droplets is plotted as a function of the dimensionless time $t^{*}$. Interestingly, this was the case for all droplets. However, it was found that droplets with lower contact angles needed more time and tended to move around in a lower range than droplets with higher contact angles. Additionally, in \cref{fig:singleHeatSource} the droplet-motion is shown for incremental increases in the heat-flux from $Q_s = 0.2$ to $Q_s = 1.0$. This indicates that there are configurations in which droplets with lower contact angle can pass through the heat source without displacement in the $z$-plane, while droplets with higher contact angle move around the heat source in the $z$-plane. Most prominently, this situation can be observed for $Q_s = 0.6$.

\begin{figure}
	\centering
	\subfloat[]{
		\includegraphics[trim={150 0 150 0},clip, width=0.7\linewidth]{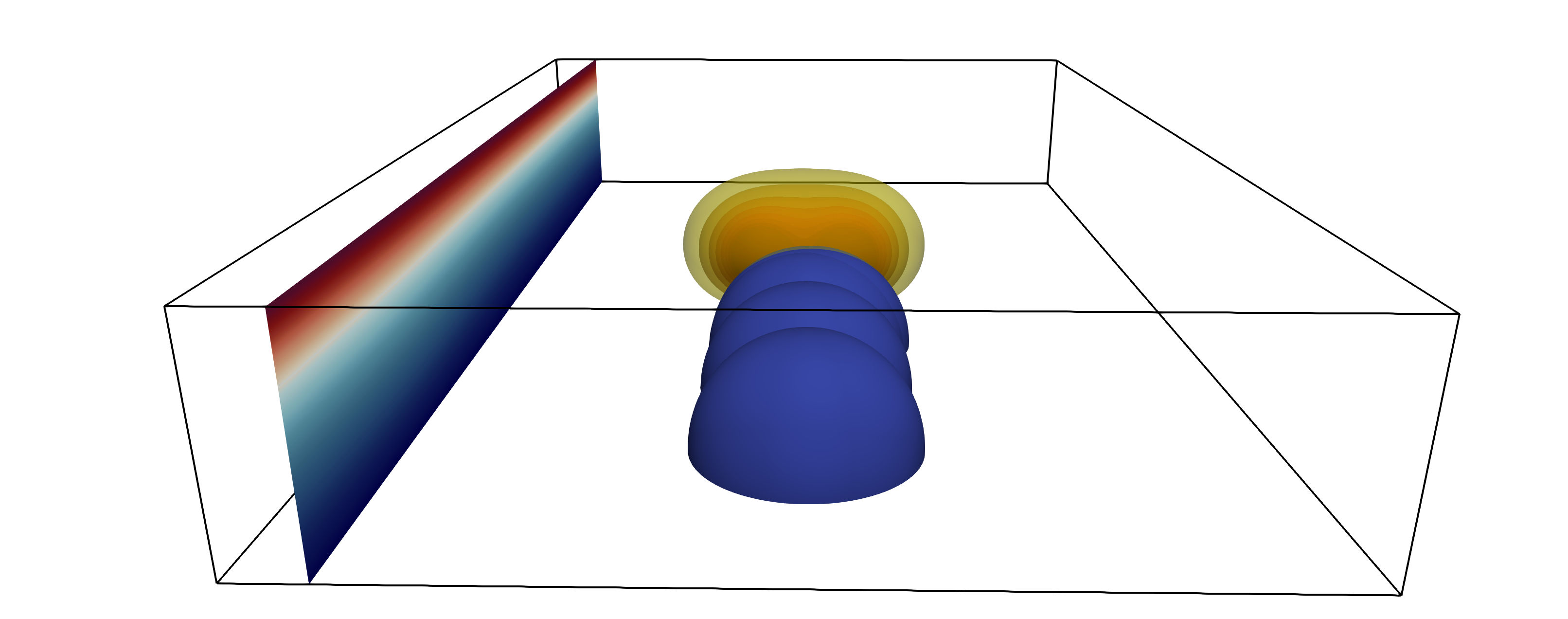}
		\label{fig:doubleHeatSourceParaview}
	}
	
	\subfloat[]{
	\input{figures/droplet_migration/3D/doubleHeatSource1_heat_flux_60.tex}
	\label{fig:doubleHeatSourcePlot}
	}
\caption{Droplet migration in a three-dimensional channel flow with various contact angles showing (a) the evolution of a droplet with $90^\circ$ contact angle where each droplet in the figure represents the movement during a period of $t^* = 10$ and (b) displacement of the droplets in the $x$- and $x$-directions with various contact angles. For the laser point, a maximal heat flux of $Q_s = 1$ was used.}
\label{fig:doubleHeatSource}
\end{figure}

Finally, flow configurations with two laser points were studied. In the first setup, the two heat sources were shifted by $\pm \nicefrac{1}{2}R$ in the $z$-direction. The three-dimensional evolution of a droplet with contact angle $\alpha = 90^\circ$ can be seen in, \cref{fig:doubleHeatSourceParaview}, while \cref{fig:doubleHeatSourcePlot} shows the droplet trajectory in the $x$- and $z$-directions with various contact angles. In this case, the blocking of the droplets can be observed for an extended period of time. However, with the applied magnitude of heat flux, the droplets with high contact angle are observed to migrate transversely and pass both laser points. Furthermore, the droplet with contact angle $\alpha = 45^\circ$ passes through the two heat sources and thus shows a very similar behaviour to that observed in the two-dimensional case, where droplets with low contact angle could slip underneath the laser point. These test cases not only provide understanding of the migratory behaviour of a droplet in shear, but showcase how the introduced modelling capability can be applied to design effective droplet capture and or manipulation devices. It is clearly shown that a three-dimensional setup shows a larger spectrum of behaviour. 

In \ref{sec:AppendixDroplet3D}, all the obtained results are presented next to each other. In this way, the effect of the increasing heat flux is shown best. \cref{fig:singleHeatSource} pictures simulations with a single heat source in the channel. Starting from a heat flux with, $Q_s = 0.2$ it can be observed that it is not possible to block any of the analysed droplets because they pass through the heat source. With increasing heat flux it stays impossible to show droplet blocking, however, for the reason that the droplets start to move around the heat source in $z$-direction. The transition can be observed best at $Q_s = 0.6$ where droplets with high contact angle move around the heat source while droplets with low contact angle pass underneath it. Next in \cref{fig:doubleHeatSource1} simulations with two heat sources are presented where the heat sources are shifted by $\pm \nicefrac{1}{2}R$. In this configuration, it is possible to show droplet blockage for simulations with higher heat flux ($Q_s \geq 0.6$). However, a secondary effect can be noticed. The heat sources are rather close to each other, and thus droplets with higher contact angle are still able to pass by on the $z$-plane. This can be seen especially well for $Q_s = 1$ where droplets with contact angle $\alpha > 90^\circ$ pass rather quickly, while droplets with lower contact angle stay blocked. Finally, in \cref{fig:doubleHeatSource2} the heat source is shifted further apart to $\pm \nicefrac{2}{3}R$. Again with lower heat flux ($Q_s \leq 0.6$) all droplets slip through the laser points. However, in none of the configurations, droplets move around the heat sources in the $z$-direction. With $Q_s = 1$ all the analysed droplets show blocking throughout the complete simulation time.

\section{Performance analysis}
\label{sec:performance}
This section shows a performance analysis of the implemented model. First, by giving detailed information about the algorithm is realised. Following that, single GPU performance results are presented and afterwards large scaling experiments are conducted.

\subsection{Algorithm}
\label{sec:algorithm}

First, the algorithm for solving thermocapillary flows using a purely \ac{LBM}-based discretisation is detailed. The algorithm starts by initialising all fields. This concerns the three PDF fields for the phase-field, the hydrodynamic and the thermal distributions, as well as the three fields for the variables of interest. These are the phase-field itself, the velocity, and the temperature field. While the phase-field and the velocity field are usually initialised statically, this means by imposing a specific value, the temperature field is often times initialised by executing the thermal solver until a certain error is reached \cite{mitchell2021computational}. After initialisation, the required ghost layer information for the thermal PDFs and the temperature field are sent to the neighbouring processes. While the data is sent using asynchronous MPI calls, the phase-field PDFs are updated and then the processes wait for the messages. As a next step, the phase-field PDFs are communicated. While this communication step happens, the hydrodynamic PDFs get updated and then the processes wait to receive the messages. The last part of the algorithm is the communication of the hydrodynamic PDFs and the phase-field. This communication happens during the update of the thermal PDFs. It is important to note here that the update of the PDFs includes the stream-collide algorithm, as well as the boundary conditions. By applying this strategy, we are able to hide all communications by overlapping them with computations.

For all stream-collide sequences in the algorithm, a one step, two grid algorithm is used by applying a pull pattern to the PDF values. For simple flows, a completely local collision has been previously proposed for the phase-field \ac{LBM}. This means it does not rely on finite differences to calculate the curvature of the phase field~\cite{Geier2015}. This allows the application of more advanced streaming patterns like the AA-pattern or the EsoTwist \cite{Wittmann2016, Geier2017EsoTwist, Lehmann2022Streaming}. However, the algorithm was shown to be less accurate than finite differences, and it did not include the temperature field needed to calculate the surface tension. As such, these advanced streaming patterns are not applied to optimise performance in this study. \par

\IncMargin{.35em}
\begin{algorithm}[htb]
	Initialisation of all fields\;
	\For{each time step t}{
		Start communication of thermal PDFs $\boldsymbol{f}$ and temperature $T$\;
		Update phase-field PDFs $\boldsymbol{h}$\;
		Wait for the communication to finish\;
		\BlankLine
		Start communication of phase-field PDFs $\boldsymbol{h}$\;
		Update hydrodynamic PDFs $\boldsymbol{g}$\;
		Wait for the communication to finish\;
		\BlankLine
		Start communication of hydrodynamic PDFs $\boldsymbol{g}$ and phase-field $\phi$\;
		Update thermal PDFs $\boldsymbol{f}$\;
		Wait for the communication to finish\;
	}
	\caption{Thermocapillary algorithm using an \ac{LBM} solver for the heat equation.}
	\label{alg:thermocapillaryLBM}
\end{algorithm}
\DecMargin{.35em}

In comparison to Algorithm~\ref{alg:thermocapillaryLBM}, Algorithm~\ref{alg:thermocapillaryRK} solves the same underlying partial differential equations, however, it uses a \ac{RK} solver instead of a \ac{LBM} scheme to solve \cref{eqn-temperature}. It starts similarly by initialising the PDF fields and the fields of interest. As a next step, the hydrodynamic PDFs are communicated with the temperature field. While the communication happens, the phase-field PDFs are updated. In the next step, the phase-field PDFs are communicated while updating the hydrodynamic PDFs. A difference of using an \ac{RK} scheme for solving \cref{eqn-temperature} is that the phase-field is needed to solve the equation. This is not the case when using an \ac{LBM} scheme, as shown in Algorithm~\ref{alg:thermocapillaryLBM}. Thus, the phase-field needs to be communicated and can not be directly overlapped with computation. However, since the phase-field is only a scalar field, the communication overhead is relatively small in comparison to the PDF fields. To finish the algorithm, the \ac{RK} scheme is applied to update the temperature field.

\IncMargin{.35em}
\begin{algorithm}[htb]
	Initialisation of all fields\;
	\For{each time step t}{
		Start communication of hydrodynamic PDFs $\boldsymbol{g}$ and temperature $T$\;
		Update phase-field PDFs $\boldsymbol{h}$\;
		Wait for the communication to finish\;
		\BlankLine
		Start communication of phase-field PDFs $\boldsymbol{h}$\;
		Update hydrodynamic PDFs $\boldsymbol{g}$\;
		Wait for the communication to finish\;
		\BlankLine
		Communication of phase-field $\phi$ \;
		\BlankLine
		Perform Runge-Kutta temperature update\;
	}
	\caption{Thermocapillary algorithm using an \ac{RK} solver for the heat equation.}
	\label{alg:thermocapillaryRK}
\end{algorithm}
\DecMargin{.35em}

Before discussing the performance of the two algorithms in detail, it must be noted that the \ac{RK} scheme consumes less memory than the LBM scheme. For the LBM solver, $2 * Q$ PDF values per cell are needed. Since \cref{eqn-temperature} can be solved using a minimal D3Q7 stencil this equates only to $14$ floating-point values. However, it is still seven times more than the \ac{RK} solver needs with only two floating-point values per cell. Nevertheless, it must be put in the perspective that the hydrodynamic and the phase-field solver consume by far the most amount of memory, and so the overall impact on memory is relatively low.

\subsection{Performance on a single GPU}
\label{sec:performanceSingleGPU}
For performance analysis of the thermocapillary flow model, the NVIDIA A100 and the AMD MI250 were used due to their prevalence in modern, large-scale supercomputers\footnote{\url{https://www.top500.org/lists/top500/list/2023/06/}}. At this day, the AMD MI250 GPU even powers the first exascale system, namely the Frontier\footnote{\url{https://www.olcf.ornl.gov/frontier/}} supercomputer at the Oak Ridge National lab. Furthermore, enabling computations on GPU has a significant advantage in the \ac{LBM} due to the increased memory bandwidth provided by the hardware~\cite{Holzer2021, lbmpy}. In this section, the performance of the model on a single GPU is investigated. These findings were then extended in \cref{sec:scaling} with a scaling experiment.

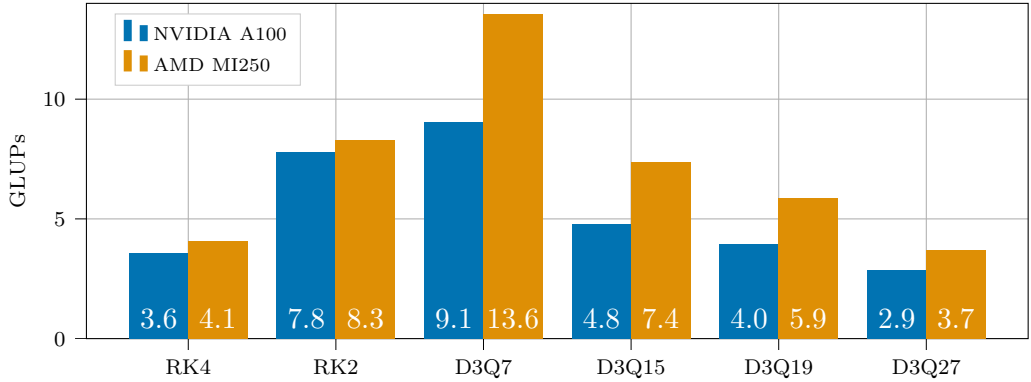
\begin{figure}[htb]
	\centering
\begin{tikzpicture}
	
	\definecolor{darkgray176}{RGB}{176,176,176}
	\definecolor{steelblue52127178}{RGB}{52,127,178}
	
	\begin{axis}[
		height=6cm,
		tick align=outside,
		tick pos=left,
		legend cell align={left},
		legend style={at={(0.03,0.97)}, anchor=north west, draw=white!80.0!black},
		width=14cm,
		x grid style={darkgray176},
		xmajorgrids,
		xmin=-0.69, xmax=5.69,
		xtick style={color=black},
		xtick={0,1,2,3,4,5},
		xticklabels={RK4,RK2,D3Q7,D3Q15,D3Q19,D3Q27},
		y grid style={darkgray176},
		ylabel={GLUPs},
		ymajorgrids,
		ymin=0, ymax=14.000,
		ytick style={color=black}
		]
		\draw[draw=none,fill=color1] (axis cs:-0.4,0) rectangle (axis cs:0,3.573);
		\draw[draw=none,fill=color2] (axis cs:0,0) rectangle (axis cs:0.4,4.060);
		
		\draw[draw=none,fill=color1] (axis cs:0.6,0) rectangle (axis cs:1.0,7.781);
		\draw[draw=none,fill=color2] (axis cs:1.0,0) rectangle (axis cs:1.4,8.308);
		
		\draw[draw=none,fill=color1] (axis cs:1.6,0) rectangle (axis cs:2.0,9.058);
		\draw[draw=none,fill=color2] (axis cs:2.0,0) rectangle (axis cs:2.4,13.544);
		
		\draw[draw=none,fill=color1] (axis cs:2.6,0) rectangle (axis cs:3,4.785);
		\draw[draw=none,fill=color2] (axis cs:3,0) rectangle (axis cs:3.4,7.352);
		
		\draw[draw=none,fill=color1] (axis cs:3.6,0) rectangle (axis cs:4,3.949);
		\draw[draw=none,fill=color2] (axis cs:4,0) rectangle (axis cs:4.4,5.850);
		
		\draw[draw=none,fill=color1] (axis cs:4.6,0) rectangle (axis cs:5,2.859);
		\draw[draw=none,fill=color2] (axis cs:5,0) rectangle (axis cs:5.4,3.712);
		\draw (axis cs:-0.2,0) node[
		scale=1.1,
		anchor=south,
		text=white,
		rotate=0.0
		]{3.6};
		\draw (axis cs:0.2,0) node[
		scale=1.1,
		anchor=south,
		text=white,
		rotate=0.0
		]{4.1};
		\draw (axis cs:0.8,0) node[
		scale=1.1,
		anchor=south,
		text=white,
		rotate=0.0
		]{7.8};
		\draw (axis cs:1.2,0) node[
		scale=1.1,
		anchor=south,
		text=white,
		rotate=0.0
		]{8.3};
		\draw (axis cs:1.8,0) node[
		scale=1.1,
		anchor=south,
		text=white,
		rotate=0.0
		]{9.1};
		\draw (axis cs:2.2,0) node[
		scale=1.1,
		anchor=south,
		text=white,
		rotate=0.0
		]{13.6};
		\draw (axis cs:2.8,0) node[
		scale=1.1,
		anchor=south,
		text=white,
		rotate=0.0
		]{4.8};
		\draw (axis cs:3.2,0) node[
		scale=1.1,
		anchor=south,
		text=white,
		rotate=0.0
		]{7.4};
		\draw (axis cs:3.8,0) node[
		scale=1.1,
		anchor=south,
		text=white,
		rotate=0.0
		]{4.0};
		\draw (axis cs:4.2,0) node[
		scale=1.1,
		anchor=south,
		text=white,
		rotate=0.0
		]{5.9};
		\draw (axis cs:4.8,0) node[
		scale=1.1,
		anchor=south,
		text=white,
		rotate=0.0
		]{2.9};
		\draw (axis cs:5.2,0) node[
		scale=1.1,
		anchor=south,
		text=white,
		rotate=0.0
		]{3.7};
		
\addlegendimage{ybar,ybar legend,fill=color1,draw opacity=0};
\addlegendentry{\scriptsize{NVIDIA A100}}
\addlegendimage{ybar,ybar legend,fill=color2,draw opacity=0};
\addlegendentry{\scriptsize{AMD MI250}}
	\end{axis}
	
\end{tikzpicture}
	\caption{Comparison of different LBM schemes and \ac{RK} schemes to solve the evolution of the temperature field.}
	\label{fig:comparissonrklbm}
\end{figure}

As shown in \cref{sec:algorithm}, it is common in the literature to use either a \ac{RK} scheme or a \ac{LBM} to solve the evolution of the temperature field. Thus, as a first step, the two possibilities should be compared with each other. For the \ac{LBM} discretisation the solver is analysed using a D3Q7, D3Q15, D3Q19 and a D3Q27 stencil (see \cref{sec:stencils}). While hydrodynamic \ac{LBM} solvers often rely on a D3Q19 or D3Q27 stencil, it is common for advection-diffusion systems to employ a much smaller stencil, namely, the D3Q7 stencil \cite{lbm_book}. For the \ac{LBM} kernel, the velocity field updated by the hydrodynamic LBM solver functions as an input for the equilibrium function and the temperature field needs to be stored/updated in every cell. This results in a memory throughput of $n_b = (Q*2 + 4)\cdot 8$ bytes per cell. For a D3Q7 stencil, $n_b = 144$ bytes per cell are required. Furthermore, to update a single cell, $n_f = 474$ FLOPs are needed, resulting in a code balance of $B_c = \frac{n_f}{n_b} = 3.3 \frac{\mathrm{FLOP}}{\mathrm{Byte}}$. This is about half of the machine balance of the A100 GPU which is $B_m = \frac{p_\mathrm{peak}}{b_s} = \frac{9.7}{1.555} = 6.2 \frac{\mathrm{FLOP}}{\mathrm{Byte}}$. Due to this, it is expected that the LBM algorithm is bound by the memory of the GPU and the maximum expected performance is $P_\mathrm{max} = \frac{b_s}{n_b} = \frac{1555 \mathrm{GB/s}}{144 \mathrm{B/cell}} = 10 \, 798$~MLUPs. This result can be confirmed by the measurement shown in \cref{fig:comparissonrklbm}, where the LBM kernel could achieve around $90 \%$ of the machine bandwidth. It must be noted that achieving higher bandwidth results is not practically possible, and even kernels like the STREAM benchmark will not achieve higher bandwidth saturation \cite{ERNST2023152, Holzer2021}. 

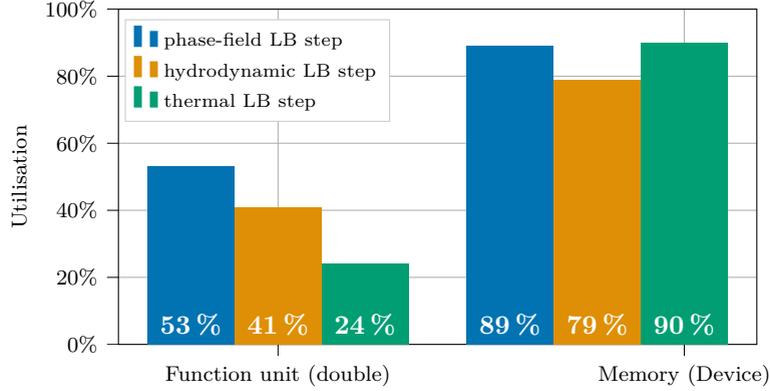
\begin{figure}[htb]
	\centering
\begin{tikzpicture}[scale=1]
\tikzset{font=\small}

\begin{axis}[
height=6cm,
width=10cm,
legend cell align={left},
legend style={at={(0.01,0.97)}, anchor=north west, draw=white!80.0!black},
tick align=outside,
tick pos=left,
x grid style={white!69.01960784313725!black},
xmajorgrids,
xmin=-1.1, xmax=1.1,
xtick style={color=black},
xtick={-0.55, 0.85},
xticklabels={Function unit (double), Memory (Device)},
y grid style={white!69.01960784313725!black},
ylabel={Utilisation},
ymajorgrids,
yticklabel={\pgfmathparse{\tick}\pgfmathprintnumber{\pgfmathresult}\%},
ymin=0, ymax=100,
ytick style={color=black}
]
\draw[fill=color1,draw opacity=0] (axis cs:-1.0,0) rectangle (axis cs:-0.7,53);
\draw[fill=color2,draw opacity=0] (axis cs:-0.7,0) rectangle (axis cs:-0.4,41);
\draw[fill=color3,draw opacity=0] (axis cs:-0.4,0) rectangle (axis cs:-0.1,24);

\draw[fill=color1,draw opacity=0] (axis cs:0.1,0) rectangle (axis cs:0.4,89);
\draw[fill=color2,draw opacity=0] (axis cs:0.4,0) rectangle (axis cs:0.7,79);
\draw[fill=color3,draw opacity=0] (axis cs:0.7,0) rectangle (axis cs:1.0,90);

\node at (axis cs:-0.85,0)[
  scale=1.0,
  anchor=south,
  text=white,
  font=\bfseries,
  rotate=0.0
]{53\,\%};
\node at (axis cs:-0.55,0)[
  scale=1.0,
  anchor=south,
  text=white,
  font=\bfseries,
  rotate=0.0
]{41\,\%};
\node at (axis cs:-0.25,0)[
  scale=1.0,
  anchor=south,
  text=white,
  font=\bfseries,
  rotate=0.0
]{24\,\%};

\node at (axis cs:0.25,0)[
  scale=1.0,
  anchor=south,
  text=white,
  font=\bfseries,
  rotate=0.0
]{89\,\%};
\node at (axis cs:0.55,0)[
scale=1.0,
anchor=south,
text=white,
font=\bfseries,
rotate=0.0
]{79\,\%};
\node at (axis cs:0.85,0)[
scale=1.0,
anchor=south,
text=white,
font=\bfseries,
rotate=0.0
]{90\,\%};

\addlegendimage{ybar,ybar legend,fill=color1,draw opacity=0};
\addlegendentry{\scriptsize{phase-field LB step}}
\addlegendimage{ybar,ybar legend,fill=color2,draw opacity=0};
\addlegendentry{\scriptsize{hydrodynamic LB step}}
\addlegendimage{ybar,ybar legend,fill=color3,draw opacity=0};
\addlegendentry{\scriptsize{thermal LB step}}
\end{axis}

\end{tikzpicture}
	\caption{Compute units utilization and memory transfer measured with Nsight Compute. The measurements are conducted on an NVIDIA A100 GPU for all \ac{LBM} solver steps individually.}
	\label{fig:profilingLBM}
\end{figure}

Similarly, the machine balance of the AMD MI250 GPU is given by $B_m = \frac{p_\mathrm{peak}}{b_s} = \frac{45.3}{3.2} = 14.2 \frac{\mathrm{FLOP}}{\mathrm{Byte}}$. Clearly this GPU is less balanced than the NVIDIA A100 with a strong favour on the compute unit part. Furthermore, as shown by Lehmann et al.~\cite{Lehmann2022Performance} it is not possible to reach more than about $50$~\% of the bandwidth with lattice Boltzmann kernels. This can be confirmed in \cref{fig:comparissonrklbm}. Although an AMD MI250 is listed with about two times the memory bandwidth than an NVIDIA A100 it was not possible to reach much more than $1.5$ of speedup.
The same experiment was conducted using stencils with a higher number of discrete velocities. These kernels saturate the bandwidth based on the same argument. However, since more data reads and writes are required, the overall performance decreases. Conducting profiling results utilising Nsight Compute\footnote{\url{https://developer.nvidia.com/nsight-compute}} shows that all \ac{LBM} kernels are limited by the bandwidth of the hardware. However, the hydrodynamic solver cannot exploit the memory bandwidth as well as the other compute kernel due to high pressure on the cache caused by the several finite difference stencils incorporated in the force term of the LBM scheme. This is summarised in \cref{fig:profilingLBM}.

\begin{figure}[htb]
	\centering
\begin{tikzpicture}[scale=1]
\tikzset{font=\small}

\begin{axis}[
height=6cm,
width=10cm,
legend cell align={left},
legend style={at={(0.01,0.97)}, anchor=north west, draw=white!80.0!black},
tick align=outside,
tick pos=left,
x grid style={white!69.01960784313725!black},
xmajorgrids,
xmin=-0.8, xmax=0.8,
xtick style={color=black},
xtick={-0.4, 0.4},
xticklabels={Function unit (double), Memory (Device)},
y grid style={white!69.01960784313725!black},
ylabel={Utilisation},
ymajorgrids,
yticklabel={\pgfmathparse{\tick}\pgfmathprintnumber{\pgfmathresult}\%},
ymin=0, ymax=100,
ytick style={color=black}
]
\draw[fill=color1,draw opacity=0] (axis cs:-0.7,0) rectangle (axis cs:-0.4,65);
\draw[fill=color2,draw opacity=0] (axis cs:-0.4,0) rectangle (axis cs:-0.1,57);
\draw[fill=color1,draw opacity=0] (axis cs:0.1,0) rectangle (axis cs:0.4,54);
\draw[fill=color2,draw opacity=0] (axis cs:0.4,0) rectangle (axis cs:0.7,50);

\node at (axis cs:-0.55,0)[
  scale=1.0,
  anchor=south,
  text=white,
  font=\bfseries,
  rotate=0.0
]{65\,\%};
\node at (axis cs:-0.25,0)[
  scale=1.0,
  anchor=south,
  text=white,
  font=\bfseries,
  rotate=0.0
]{57\,\%};

\node at (axis cs:0.25,0)[
scale=1.0,
anchor=south,
text=white,
font=\bfseries,
rotate=0.0
]{54\,\%};
\node at (axis cs:0.55,0)[
scale=1.0,
anchor=south,
text=white,
font=\bfseries,
rotate=0.0
]{50\,\%};

\addlegendimage{ybar,ybar legend,fill=color1,draw opacity=0};
\addlegendentry{\scriptsize{RK stage 1}}
\addlegendimage{ybar,ybar legend,fill=color2,draw opacity=0};
\addlegendentry{\scriptsize{RK stage 2}}
f\end{axis}

\end{tikzpicture}
	\caption{Compute units utilization and memory transfer measured with Nsight Compute. The measurements are conducted on a NVIDIA A100 GPU for all steps of the second-order \ac{RK} scheme used to solve the temperature equation.}
	\label{fig:profilingRK}
\end{figure}
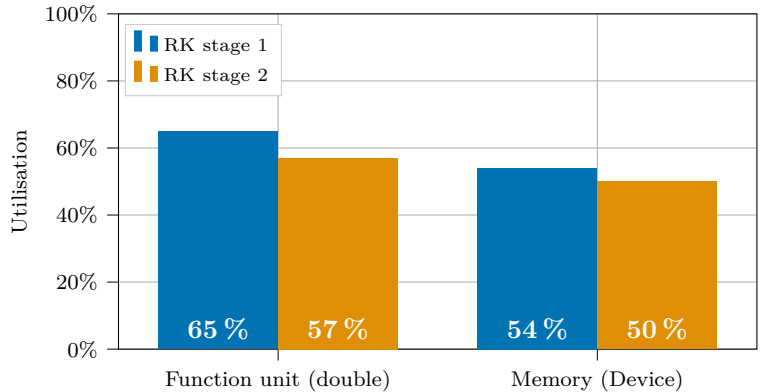

On the other side, using an \ac{RK} scheme to resolve the evolution of the temperature field comes with some flexibility as well. Oftentimes in the literature, a fourth-order \ac{RK} scheme is used~\cite{mitchell2021computational}. However, as shown earlier, a second-order \ac{RK} scheme is also feasible when considering that the \ac{RK} scheme is employed alongside the \ac{LBM} solvers for the hydrodynamics and the phase-field. In the scope of this work, the \ac{RK} scheme is implemented as consecutive compute kernels for \cref{eqn-rk2}. Each of these kernels compute the gradient and the Laplacian of a previous field, along with the gradient of the phase-field. With these, an update rule is employed that forms the result of a temporary field until the final compute kernel, where the result is stored on the temperature field itself. Furthermore, the velocity field is needed to fulfil \cref{eqn-temperature}. Ideally, each field value is only read once, which would be the case when all neighbouring cell access could be cached. In this case, the code balance of each of the stages would be approximately $B_c = 3.7 \frac{\mathrm{FLOP}}{\mathrm{Byte}}$. However, looking at the profiling results conducted with Nsight Compute\footnote{\url{https://developer.nvidia.com/nsight-compute}} reveals that the ideal assumption can not be fully met, resulting in compute bound kernels. Each kernel utilises approximately \nicefrac{$2$}{$3$} of the maximum compute capacity, which is a common value for compute bound kernels \cite{ERNST2023152}. In \cref{fig:profilingRK}, profiling results for each \ac{RK} stage are shown respectively. Therein, the two stages correspond to \cref{eqn-rk2} accordingly.

\subsection{Scaling results}
\label{sec:scaling}

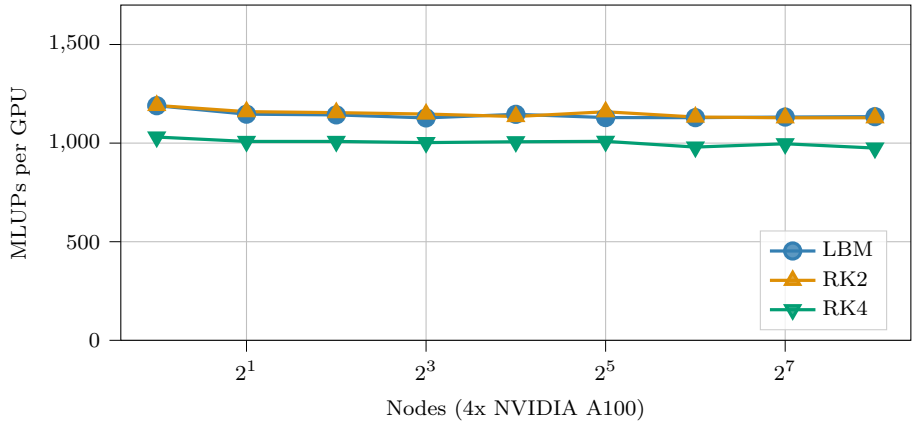
\begin{figure}[htb]
	\centering
\begin{tikzpicture}

\definecolor{darkcyan2158115}{RGB}{2,158,115}
\definecolor{darkorange2221435}{RGB}{222,143,5}
\definecolor{lightgray204}{RGB}{204,204,204}
\definecolor{steelblue52127178}{RGB}{52,127,178}

\begin{axis}[
height=6cm,
legend cell align={left},
legend style={
  fill opacity=0.8,
  draw opacity=1,
  text opacity=1,
  at={(0.97,0.03)},
  anchor=south east,
  draw=lightgray204
},
log basis x={2},
tick align=outside,
tick pos=left,
width=12cm,
xlabel={Nodes (4x NVIDIA A100)},
xmajorgrids,
xmin=0.757858283255199, xmax=337.794025157861,
xmode=log,
xtick style={color=black},
xtick={0.125,0.5,2,8,32,128,512,2048},
xticklabels={
  \(\displaystyle {2^{-3}}\),
  \(\displaystyle {2^{-1}}\),
  \(\displaystyle {2^{1}}\),
  \(\displaystyle {2^{3}}\),
  \(\displaystyle {2^{5}}\),
  \(\displaystyle {2^{7}}\),
  \(\displaystyle {2^{9}}\),
  \(\displaystyle {2^{11}}\)
},
ylabel={MLUPs per GPU},
ymajorgrids,
ymin=0, ymax=1700,
ytick style={color=black}
]
\addplot [very thick, steelblue52127178, mark=*, mark size=3, mark options={solid}]
table {%
1 1189.62055173655
2 1146.73040220635
4 1143.10417334142
8 1127.77209874831
16 1145.95410365661
32 1129.51616579905
64 1129.05303583687
128 1132.104473335
256 1134.10255306019
};
\addlegendentry{LBM}
\addplot [very thick, darkorange2221435, mark=triangle*, mark size=3, mark options={solid}]
table {%
1 1190.90210672031
2 1159.47095046949
4 1154.46004778641
8 1147.3565762814
16 1134.82453009111
32 1158.95573503371
64 1132.17305128519
128 1128.31350482707
256 1128.45427594165
};
\addlegendentry{RK2}
\addplot [very thick, darkcyan2158115, mark=triangle*, mark size=3, mark options={solid,rotate=180}]
table {%
1 1030.80267046178
2 1008.40636675412
4 1008.45225007546
8 1002.5824285448
16 1006.83484047906
32 1008.89438897557
64 979.870373767468
128 996.618054683031
256 974.859402423712
};
\addlegendentry{RK4}
\end{axis}

\end{tikzpicture}
	\caption{Weak scaling performance on the JUWELS Booster supercomputer. The scaling results are conducted with a D3Q7 LBM solver for the evolution of the temperature field as well as a second- and fourth-order Runge-Kutta scheme.}
	\label{fig:weakscalingjuwelsbooster}
\end{figure}

Building on \cref{sec:performanceSingleGPU}, the single GPU results are extended in a scaling experiment conducted on the JUWELS Booster supercomputer as well as the GPU partition of the LUMI supercomputer (LUMI-G). For this experiment, the droplet scenario from \cref{sec:3d-laser} is put in larger and larger domains to form a weak scaling scenario. The results for the NVIDIA-based system are summarised in \cref{fig:weakscalingjuwelsbooster}. On each A100 GPU, a domain size of $512 \times 256 \times 256$ cells was used. This is scaled up to $256$ nodes containing $1024$ GPUs in total. On $1024$ GPUs a system of almost 35 billion grid cells was solved. At this scale, it is still possible to obtain about $30$ timesteps per second. With the total performance of about one TLUPs $1 \cdot 10^12$ grid cells can be updated per second.

As a second benchmark, the same experiment was repeated on the LUMI supercomputer. Each node of the GPU partition contains four AMD MI250 GPUs. The weak scaling experiment was conducted on up to $4096$ GPUs. Each AMD MI250 contains two graphics compute dies (GCD). Each of those is operated with an individual MPI Rank and thus one block of the domain decomposition. Thus, while using four times more GPUs than on the JUWELS Booster system, eight times more MPI ranks are used and the domain size is eight times bigger. This results in a domain of almost $275$ billion grid nodes. On $4096$ GPUs a total performance of more than four TLUPs could have been achieved.  

\begin{figure}[htb]
	\centering
\begin{tikzpicture}

\definecolor{darkcyan2158115}{RGB}{2,158,115}
\definecolor{darkorange2221435}{RGB}{222,143,5}
\definecolor{lightgray204}{RGB}{204,204,204}
\definecolor{steelblue52127178}{RGB}{52,127,178}

\begin{axis}[
height=6cm,
legend cell align={left},
legend style={
  fill opacity=0.8,
  draw opacity=1,
  text opacity=1,
  at={(0.97,0.03)},
  anchor=south east,
  draw=lightgray204
},
log basis x={2},
tick align=outside,
tick pos=left,
width=12cm,
xlabel={Nodes (4x AMD MI250)},
xmajorgrids,
xmin=0.707106781186548, xmax=1448.15468787005,
xmode=log,
xtick style={color=black},
xtick={0.125,0.5,2,8,32,128,512,2048,8192},
xticklabels={
  \(\displaystyle {2^{-3}}\),
  \(\displaystyle {2^{-1}}\),
  \(\displaystyle {2^{1}}\),
  \(\displaystyle {2^{3}}\),
  \(\displaystyle {2^{5}}\),
  \(\displaystyle {2^{7}}\),
  \(\displaystyle {2^{9}}\),
  \(\displaystyle {2^{11}}\),
  \(\displaystyle {2^{13}}\)
},
ylabel={MLUPs per GPU},
ymajorgrids,
ymin=0, ymax=1700,
ytick style={color=black}
]
\addplot [very thick, steelblue52127178, mark=*, mark size=3, mark options={solid}]
table {%
1 1626.67781137688
2 1629.04205727501
4 1611.78749558956
8 1581.70574573045
16 1585.55852341161
32 1579.33926527911
64 1574.72068718895
128 1567.06726495537
256 1567.70693711369
512 1558.14617103797
1024 1550.18651371302
};
\addlegendentry{LBM}
\addplot [very thick, darkorange2221435, mark=triangle*, mark size=3, mark options={solid}]
table {%
1 1472.48816286642
2 1475.78628500221
4 1459.89133558456
8 1425.95854570432
16 1435.53183563702
32 1434.4828072305
64 1420.41622203345
128 1411.34704925606
256 1395.28633206522
512 1357.60514675355
1024 1274.16466310335
};
\addlegendentry{RK2}
\addplot [very thick, darkcyan2158115, mark=triangle*, mark size=3, mark options={solid,rotate=180}]
table {%
1 1244.69928403598
2 1246.12577245374
4 1232.44572733134
8 1208.59793014961
16 1214.3988332048
32 1199.82000124151
64 1188.54482413929
128 1193.06376558403
256 1164.21092161941
512 1115.56668041792
1024 1053.04231032753
};
\addlegendentry{RK4}
\end{axis}

\end{tikzpicture}
	\caption{Weak scaling performance on the LUMI supercomputer on up to $4096$ AMD MI250 GPUs. The scaling results are conducted with a D3Q7 LBM solver for the evolution of the temperature field as well as a second- and fourth-order Runge-Kutta scheme.}
	\label{fig:weakscalinglumi}
\end{figure}
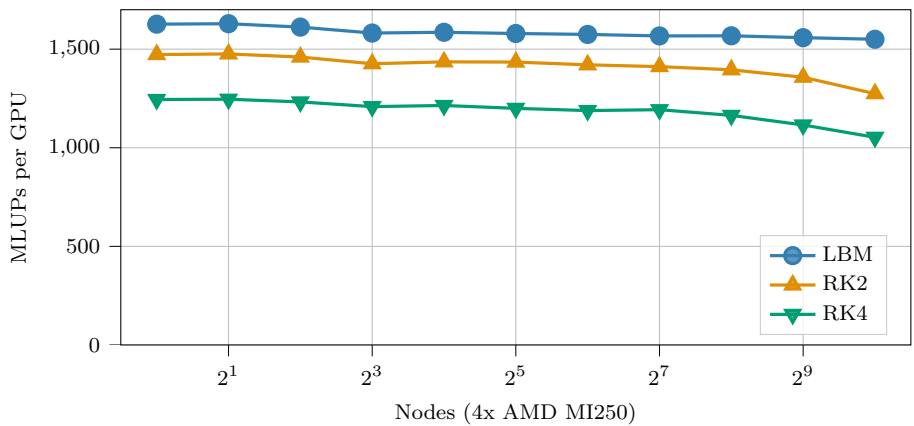

In summary, it could have been shown that both versions, either with \ac{LBM} or with the \ac{RK} scheme, have very good scaling behaviour since the largest portion of the communication can be hidden behind computations. However, the \ac{RK}4 scheme comes with a slightly lower overall performance when compared with the \ac{RK}2 and \ac{LBM} schemes, which achieve very similar performance. This was found on both NVIDIA- and AMD-based systems. On the AMD-based system, even the \ac{RK}2 scheme shows noticeable performance degeneration.

\section{Conclusions}
This study developed a new thermocapillary, central-moment phase-field \ac{LBM} and presented a highly-optimised implementation in the open-source framework, \walberla. It consolidated the approaches commonly observed in the literature for resolving the energy equation and contrasted them in terms of accuracy and computational performance. It was observed that using a D3Q7 lattice stencil provided the fastest compute with minimal impact on accuracy. However, when combining the algorithm with the phase-field and hydrodynamics, the \ac{RK2} method provided very similar performance while leaving a lower memory footprint. The scalability of the algorithm was demonstrated with excellent results based on the ability to hide communication with required computation. The performance investigation was done on state-of-the-art NVIDIA and AMD GPUs and on the biggest systems in Europe at the time of writing.

The capability that this model and implementation provide was applied to study the ability of local heating (e.g. a source analogous to laser tweezers) to stop the shear migration of droplets. It was observed that a 2D representation of this inherently 3D scenario was not sufficient to determine when capture would occur. Using only a single heated source in the three-dimensional test case leads to the observation that the droplets circumnavigated the hot spot, allowing for continued migration. With two heated sources, the droplet was able to be stopped, however, a significant influence on the placement of the heat sources and the contact angle of the droplet was observed. The model and implementation presented here are expected to facilitate future research on the design of microfluidic devices reliant on droplet manipulation and control.

\FloatBarrier
\newpage
\appendix
\section{Analytic solution of planar heated channel} 
\label{sec:anlytical-solution}

The analytical solution for the first benchmark problem is given by \cite{pendse_esmaeeli_2010},
\begin{small}
\begin{eqnarray}
	u(x,y)  &=& \begin{cases}
		U_{max}\{ [ {C_1}^b + \omega({C_2}^b +{C_3}^b y ) ]\cosh(\omega y)+ ( {C_3}^b + \omega {C_1}^b y )\sinh(\omega y)   \}\times \sin(\omega x), \ \ \ y\leq 0, \\
		U_{max}\{ [ {C_1}^a + \omega({C_2}^a +{C_3}^a y ) ]\cosh(\omega y)+ ( {C_3}^a + \omega {C_1}^a y )\sinh(\omega y)   \}\times \sin(\omega x), \ \ \  y > 0, \\
	\end{cases} \\
	v(x,y)  &=& \begin{cases}
		-\omega U_{max} [ {C_1}^b y\cosh(\omega y) + ({C_2}^b  +{C_3}^b y)
		\sinh(\omega y)]\times \cos(\omega x), \ \ \  y\leq 0, \\
		-\omega U_{max} [ {C_1}^a y\cosh(\omega y) + ({C_2}^a  +{C_3}^a y)
		\sinh(\omega y)]\times \cos(\omega x), \ \ \  y > 0, \\
	\end{cases}\\
	T(x,y)  &=& \begin{cases}
		\frac{(T_c - T_h)y + k_r T_c b + T_h a}{a + k_r b} 
		+ T_0 f(a_r , b_r , k_r) \sinh(a_r -\omega y)\cos(\omega x),  \ \ \  y\leq 0, \\
		\frac{(T_c - T_h)y + k_r T_c b + T_h a}{a + k_r b} + T_0 f(a_r , b_r , k_r)\times  [ \sinh(a_r)\cosh(\omega y)
		-k_r \sinh(\omega y)\cosh(a_r)]\cos(\omega x), \ \ \  y > 0. \\
	\end{cases}
\end{eqnarray}
\end{small}
where the $x$- and $y$-components of velocity are given by $u$ and $v$, respectively. The additional parameters in the previous equations are expressed as,
\begin{small}
	\begin{equation}
		a_r = a\omega, \ \ b_r  = b\omega,
	\end{equation}
	\begin{equation}
		f(a_r , b_r , k_r) = \frac{1}{ k_r \sinh(b_r)\cosh(a_r) + \sinh(a_r)\cosh(b_r) },
	\end{equation}
	\begin{equation}
		{C_1}^a = \frac{\sinh^2 (a_r)}{\sinh^2 (a_r) - {a_r}^2}, \ {C_1}^b = \frac{\sinh^2 (b_r)}{\sinh^2 (b_r) - {b_r}^2},
	\end{equation}
	\begin{equation}
		{C_2}^a = \frac{-aa_r}{\sinh^2 (a_r) - {a_r}^2}, \ {C_2}^b = \frac{-bb_r}{\sinh^2 (b_r) - {b_r}^2},
	\end{equation}
	\begin{equation}
		{C_3}^a = \frac{2a_r -\sinh (2a_r)}{2[\sinh^2 (a_r) - {a_r}^2]}, \ {C_3}^b =     \frac{-2b_r +\sinh (2b_r)}{2[\sinh^2 (b_r) - {b_r}^2]},
	\end{equation}
	\begin{equation}
		U_{max} = -\left(\frac{T_0 \sigma_T}{\mu_h}g(a_r,b_r,k_r)h(a_r,b_r,k_r)\right),
	\end{equation}
	\begin{equation}
		g(a_r,b_r,k_r) = \sinh a_r \times f(a_r,b_r,k_r),
	\end{equation}
	\begin{equation}
		\begin{split}
			h(a_r,b_r,k_r) = \{[\sinh^2 (a_r) - {a_r}^2]\times [\sinh^2 (b_r) - {b_r}^2]    \}\\
			\times \{ \mu_r [\sinh^2 (b_r) - {b_r}^2]\times [\sinh (2a_r) - {2a_r}] \\
			+ [\sinh^2 (a_r) - {a_r}^2]\times[ \sinh(2b_r) - {2b_r}] \}^{-1}.
		\end{split}
	\end{equation}
\end{small}
\section{Three-dimensional LB Stencils}
\label{sec:stencils}
For the three-dimensional flow scenarios, four different lattice stencils are used and studied in terms of numerical accuracy and computational performance. These are the D3Q7, D3Q15, D3Q19 and the D3Q27 stencil. 

\begin{figure}[htb]
	\centering
	
	\subfloat[D3Q7]{
		\scalebox{.6}{\begin{tikzpicture}[scale=0.9]
	
	\def\x{0.5}
	
	\draw[thick] (0, 0, 0) rectangle (4, 4, 0);
	\draw[thick] (\x, 0, 4) rectangle (4 + \x, 4, 4);
	
	\draw[thick, dotted] (0, 2, 0) -- (0+\x, 2, 4);
	\draw[thick, dotted] (0 + \x, 2, 4) -- (4 + \x, 2, 4);
	\draw[thick, dotted] (4 + \x, 2, 4) -- (4, 2, 0);
	\draw[thick, dotted] (4, 2, 0) -- (0, 2, 0);
	
	\draw[thick, dotted] (0+\x/2, 0, 2) -- (4 +\x/2, 0, 2);
	\draw[thick, dotted] (4+\x/2, 0, 2) -- (4+\x/2, 4, 2);
	\draw[thick, dotted] (4+\x/2, 4, 2) -- (0+\x/2, 4, 2);
	\draw[thick, dotted] (0+\x/2, 4, 2) -- (0+\x/2, 0, 2);
	
	\draw[thick, dotted] (2, 0, 0) -- (2 + \x, 0, 4);
	\draw[thick, dotted] (2 + \x, 0, 4) -- (2 + \x, 4, 4);
	\draw[thick, dotted] (2 + \x, 4, 4) -- (2, 4, 0);
	\draw[thick, dotted] (2, 4, 0) -- (2, 0, 0);
	
	\draw[thick] (0, 0, 0) -- (0 + \x, 0, 4);
	\draw[thick] (4, 0, 0) -- (4 + \x, 0, 4);
	\draw[thick] (4, 4, 0) -- (4 + \x, 4, 4);
	\draw[thick] (0, 4, 0) -- (0 + \x, 4, 4);

	\draw [ultra thick, -latex, lssblue] (2+\x/2, 2, 2) -- (2+\x/2, 4, 2) node [black, above] {$1$};
	\draw [ultra thick, -latex, lssblue] (2+\x/2, 2, 2) -- (2+\x/2, 0, 2) node [black, below] {$2$};
	
	\draw [ultra thick, -latex, lssblue] (2+\x/2, 2, 2) -- (0+\x/2, 2, 2) node [black, left] {$3$};
	\draw [ultra thick, -latex, lssblue] (2+\x/2, 2, 2) -- (4+\x/2, 2, 2) node [black, right] {$4$};
	
	\draw [ultra thick, -latex, lssblue] (2+\x/2, 2, 2) -- (2 + \x, 2, 4) node [black, below] {$5$};
	\draw [ultra thick, -latex, lssblue] (2+\x/2, 2, 2) -- (2, 2, 0) node [black, above right=-0.1] {$6$};
	
	\draw [ultra thick, -latex, lssblue!80] (2+\x/2, 2, 2) -- (2 + \x, 0, 4) node [black, below] { };

	\node[white] at (2+\x/2,2,2) {$0$};
	
\end{tikzpicture}}
		\label{fig:d3q7}
	}
		\subfloat[D3Q15]{
		\scalebox{.6}{\begin{tikzpicture}[scale=0.9]

\def\x{0.5}

\draw[thick] (0, 0, 0) rectangle (4, 4, 0);
\draw[thick] (\x, 0, 4) rectangle (4 + \x, 4, 4);

\draw[thick, dotted] (0, 2, 0) -- (0+\x, 2, 4);
\draw[thick, dotted] (0 + \x, 2, 4) -- (4 + \x, 2, 4);
\draw[thick, dotted] (4 + \x, 2, 4) -- (4, 2, 0);
\draw[thick, dotted] (4, 2, 0) -- (0, 2, 0);

\draw[thick, dotted] (0+\x/2, 0, 2) -- (4 +\x/2, 0, 2);
\draw[thick, dotted] (4+\x/2, 0, 2) -- (4+\x/2, 4, 2);
\draw[thick, dotted] (4+\x/2, 4, 2) -- (0+\x/2, 4, 2);
\draw[thick, dotted] (0+\x/2, 4, 2) -- (0+\x/2, 0, 2);

\draw[thick, dotted] (2, 0, 0) -- (2 + \x, 0, 4);
\draw[thick, dotted] (2 + \x, 0, 4) -- (2 + \x, 4, 4);
\draw[thick, dotted] (2 + \x, 4, 4) -- (2, 4, 0);
\draw[thick, dotted] (2, 4, 0) -- (2, 0, 0);

\draw[thick] (0, 0, 0) -- (0 + \x, 0, 4);
\draw[thick] (4, 0, 0) -- (4 + \x, 0, 4);
\draw[thick] (4, 4, 0) -- (4 + \x, 4, 4);
\draw[thick] (0, 4, 0) -- (0 + \x, 4, 4);

\draw [ultra thick, -latex, lssblue] (2+\x/2, 2, 2) -- (2+\x/2, 4, 2) node [black, above] {$1$};
\draw [ultra thick, -latex, lssblue] (2+\x/2, 2, 2) -- (2+\x/2, 0, 2) node [black, below] {$2$};

\draw [ultra thick, -latex, lssblue] (2+\x/2, 2, 2) -- (0+\x/2, 2, 2) node [black, left] {$3$};
\draw [ultra thick, -latex, lssblue] (2+\x/2, 2, 2) -- (4+\x/2, 2, 2) node [black, right] {$4$};

\draw [ultra thick, -latex, lssblue] (2+\x/2, 2, 2) -- (2 + \x, 2, 4) node [black, below] {$5$};
\draw [ultra thick, -latex, lssblue] (2+\x/2, 2, 2) -- (2, 2, 0) node [black, above right=-0.1] {$6$};

\draw [ultra thick, -latex, lssblue!60] (2+\x/2, 2, 2) -- (4 + \x, 4, 4) node [black, right] {$7$};
\draw [ultra thick, -latex, lssblue!60] (2+\x/2, 2, 2) -- (0 + \x, 4, 4) node [black, left] {$8$};
\draw [ultra thick, -latex, lssblue!60] (2+\x/2, 2, 2) -- (4 + \x, 0, 4) node [black, below] {$9$};
\draw [ultra thick, -latex, lssblue!60] (2+\x/2, 2, 2) -- (0 + \x, 0, 4) node [black, left] {$10$};
\draw [ultra thick, -latex, lssblue!60] (2+\x/2, 2, 2) -- (4, 4, 0) node [black, right] {$11$};
\draw [ultra thick, -latex, lssblue!60] (2+\x/2, 2, 2) -- (0, 4, 0) node [black, left] {$12$};
\draw [ultra thick, -latex, lssblue!60] (2+\x/2, 2, 2) -- (4, 0, 0) node [black, right] {$13$};
\draw [ultra thick, -latex, lssblue!60] (2+\x/2, 2, 2) -- (0, 0, 0) node [black, left] {$14$};

\draw [ultra thick, -latex, lssblue!80] (2+\x/2, 2, 2) -- (2 + \x, 0, 4) node [black, below] { };

\node[white] at (2+\x/2,2,2) {$0$};

\end{tikzpicture}}
		\label{fig:d3q15}
	}
		\subfloat[D3Q19]{
		\scalebox{.6}{\begin{tikzpicture}[scale=0.9]

\def\x{0.5}

\draw[thick] (0, 0, 0) rectangle (4, 4, 0);
\draw[thick] (\x, 0, 4) rectangle (4 + \x, 4, 4);

\draw[thick, dotted] (0, 2, 0) -- (0+\x, 2, 4);
\draw[thick, dotted] (0 + \x, 2, 4) -- (4 + \x, 2, 4);
\draw[thick, dotted] (4 + \x, 2, 4) -- (4, 2, 0);
\draw[thick, dotted] (4, 2, 0) -- (0, 2, 0);

\draw[thick, dotted] (0+\x/2, 0, 2) -- (4 +\x/2, 0, 2);
\draw[thick, dotted] (4+\x/2, 0, 2) -- (4+\x/2, 4, 2);
\draw[thick, dotted] (4+\x/2, 4, 2) -- (0+\x/2, 4, 2);
\draw[thick, dotted] (0+\x/2, 4, 2) -- (0+\x/2, 0, 2);

\draw[thick, dotted] (2, 0, 0) -- (2 + \x, 0, 4);
\draw[thick, dotted] (2 + \x, 0, 4) -- (2 + \x, 4, 4);
\draw[thick, dotted] (2 + \x, 4, 4) -- (2, 4, 0);
\draw[thick, dotted] (2, 4, 0) -- (2, 0, 0);

\draw[thick] (0, 0, 0) -- (0 + \x, 0, 4);
\draw[thick] (4, 0, 0) -- (4 + \x, 0, 4);
\draw[thick] (4, 4, 0) -- (4 + \x, 4, 4);
\draw[thick] (0, 4, 0) -- (0 + \x, 4, 4);

\draw [ultra thick, -latex, lssblue] (2+\x/2, 2, 2) -- (2+\x/2, 4, 2) node [black, above] {$1$};
\draw [ultra thick, -latex, lssblue] (2+\x/2, 2, 2) -- (2+\x/2, 0, 2) node [black, below] {$2$};

\draw [ultra thick, -latex, lssblue] (2+\x/2, 2, 2) -- (0+\x/2, 2, 2) node [black, left] {$3$};
\draw [ultra thick, -latex, lssblue] (2+\x/2, 2, 2) -- (4+\x/2, 2, 2) node [black, right] {$4$};

\draw [ultra thick, -latex, lssblue] (2+\x/2, 2, 2) -- (2 + \x, 2, 4) node [black, below] {$5$};
\draw [ultra thick, -latex, lssblue] (2+\x/2, 2, 2) -- (2, 2, 0) node [black, above right=-0.1] {$6$};

\draw [ultra thick, -latex, lssblue!80] (2+\x/2, 2, 2) -- (0+\x/2, 4, 2) node [black, left] {$7$};
\draw [ultra thick, -latex, lssblue!80] (2+\x/2, 2, 2) -- (4+\x/2, 4, 2) node [black, right] {$8$};
\draw [ultra thick, -latex, lssblue!80] (2+\x/2, 2, 2) -- (0+\x/2, 0, 2) node [black, left] {$9$};
\draw [ultra thick, -latex, lssblue!80] (2+\x/2, 2, 2) -- (4+\x/2, 0, 2) node [black, below] {$10$};

\draw [ultra thick, -latex, lssblue!80] (2+\x/2, 2, 2) -- (2 + \x, 4, 4) node [black, above] {$11$};
\draw [ultra thick, -latex, lssblue!80] (2+\x/2, 2, 2) -- (2 + \x, 0, 4) node [black, below] {$12$};
\draw [ultra thick, -latex, lssblue!80] (2+\x/2, 2, 2) -- (0 + \x, 2, 4) node [black, left] {$13$};
\draw [ultra thick, -latex, lssblue!80] (2+\x/2, 2, 2) -- (4 + \x, 2, 4) node [black, right] {$14$};
\draw [ultra thick, -latex, lssblue!80] (2+\x/2, 2, 2) -- (2, 4, 0) node [black, above] {$15$};

\draw [ultra thick, -latex, lssblue!80] (2+\x/2, 2, 2) -- (2, 0, 0) node [black, below] {$16$};
\draw [ultra thick, -latex, lssblue!80] (2+\x/2, 2, 2) -- (0, 2, 0) node [black, left] {$17$};
\draw [ultra thick, -latex, lssblue!80] (2+\x/2, 2, 2) -- (4, 2, 0) node [black, right] {$18$};

\node[white] at (2+\x/2,2,2) {$0$};

\end{tikzpicture}}
		\label{fig:d3q19}
	}
		\subfloat[D3Q27]{
		\scalebox{.6}{\begin{tikzpicture}[scale=0.9]

\def\x{0.5}

\draw[thick] (0, 0, 0) rectangle (4, 4, 0);
\draw[thick] (\x, 0, 4) rectangle (4 + \x, 4, 4);

\draw[thick, dotted] (0, 2, 0) -- (0+\x, 2, 4);
\draw[thick, dotted] (0 + \x, 2, 4) -- (4 + \x, 2, 4);
\draw[thick, dotted] (4 + \x, 2, 4) -- (4, 2, 0);
\draw[thick, dotted] (4, 2, 0) -- (0, 2, 0);

\draw[thick, dotted] (0+\x/2, 0, 2) -- (4 +\x/2, 0, 2);
\draw[thick, dotted] (4+\x/2, 0, 2) -- (4+\x/2, 4, 2);
\draw[thick, dotted] (4+\x/2, 4, 2) -- (0+\x/2, 4, 2);
\draw[thick, dotted] (0+\x/2, 4, 2) -- (0+\x/2, 0, 2);

\draw[thick, dotted] (2, 0, 0) -- (2 + \x, 0, 4);
\draw[thick, dotted] (2 + \x, 0, 4) -- (2 + \x, 4, 4);
\draw[thick, dotted] (2 + \x, 4, 4) -- (2, 4, 0);
\draw[thick, dotted] (2, 4, 0) -- (2, 0, 0);

\draw[thick] (0, 0, 0) -- (0 + \x, 0, 4);
\draw[thick] (4, 0, 0) -- (4 + \x, 0, 4);
\draw[thick] (4, 4, 0) -- (4 + \x, 4, 4);
\draw[thick] (0, 4, 0) -- (0 + \x, 4, 4);

\draw [ultra thick, -latex, lssblue] (2+\x/2, 2, 2) -- (2+\x/2, 4, 2) node [black, above] {$1$};
\draw [ultra thick, -latex, lssblue] (2+\x/2, 2, 2) -- (2+\x/2, 0, 2) node [black, below] {$2$};

\draw [ultra thick, -latex, lssblue] (2+\x/2, 2, 2) -- (0+\x/2, 2, 2) node [black, left] {$3$};
\draw [ultra thick, -latex, lssblue] (2+\x/2, 2, 2) -- (4+\x/2, 2, 2) node [black, right] {$4$};

\draw [ultra thick, -latex, lssblue] (2+\x/2, 2, 2) -- (2 + \x, 2, 4) node [black, below] {$5$};
\draw [ultra thick, -latex, lssblue] (2+\x/2, 2, 2) -- (2, 2, 0) node [black, above right=-0.1] {$6$};

\draw [ultra thick, -latex, lssblue!80] (2+\x/2, 2, 2) -- (0+\x/2, 4, 2) node [black, left] {$7$};
\draw [ultra thick, -latex, lssblue!80] (2+\x/2, 2, 2) -- (4+\x/2, 4, 2) node [black, right] {$8$};
\draw [ultra thick, -latex, lssblue!80] (2+\x/2, 2, 2) -- (0+\x/2, 0, 2) node [black, left] {$9$};
\draw [ultra thick, -latex, lssblue!80] (2+\x/2, 2, 2) -- (4+\x/2, 0, 2) node [black, below] {$10$};

\draw [ultra thick, -latex, lssblue!80] (2+\x/2, 2, 2) -- (2 + \x, 4, 4) node [black, above] {$11$};
\draw [ultra thick, -latex, lssblue!80] (2+\x/2, 2, 2) -- (2 + \x, 0, 4) node [black, below] {$12$};
\draw [ultra thick, -latex, lssblue!80] (2+\x/2, 2, 2) -- (0 + \x, 2, 4) node [black, left] {$13$};
\draw [ultra thick, -latex, lssblue!80] (2+\x/2, 2, 2) -- (4 + \x, 2, 4) node [black, right] {$14$};
\draw [ultra thick, -latex, lssblue!80] (2+\x/2, 2, 2) -- (2, 4, 0) node [black, above] {$15$};

\draw [ultra thick, -latex, lssblue!80] (2+\x/2, 2, 2) -- (2, 0, 0) node [black, below] {$16$};
\draw [ultra thick, -latex, lssblue!80] (2+\x/2, 2, 2) -- (0, 2, 0) node [black, left] {$17$};
\draw [ultra thick, -latex, lssblue!80] (2+\x/2, 2, 2) -- (4, 2, 0) node [black, right] {$18$};

\draw [ultra thick, -latex, lssblue!60] (2+\x/2, 2, 2) -- (4 + \x, 4, 4) node [black, right] {$19$};
\draw [ultra thick, -latex, lssblue!60] (2+\x/2, 2, 2) -- (0 + \x, 4, 4) node [black, left] {$20$};
\draw [ultra thick, -latex, lssblue!60] (2+\x/2, 2, 2) -- (4 + \x, 0, 4) node [black, below] {$21$};
\draw [ultra thick, -latex, lssblue!60] (2+\x/2, 2, 2) -- (0 + \x, 0, 4) node [black, left] {$22$};
\draw [ultra thick, -latex, lssblue!60] (2+\x/2, 2, 2) -- (4, 4, 0) node [black, right] {$23$};
\draw [ultra thick, -latex, lssblue!60] (2+\x/2, 2, 2) -- (0, 4, 0) node [black, left] {$24$};
\draw [ultra thick, -latex, lssblue!60] (2+\x/2, 2, 2) -- (4, 0, 0) node [black, right] {$25$};
\draw [ultra thick, -latex, lssblue!60] (2+\x/2, 2, 2) -- (0, 0, 0) node [black, left] {$26$};

\node[white] at (2+\x/2,2,2) {$0$};

\end{tikzpicture}}
		\label{fig:d3q27}
	}
	\caption{The discrete velocities for all three-dimensional lattice stencils used in this work.}
	\label{fig:stencils}
\end{figure}
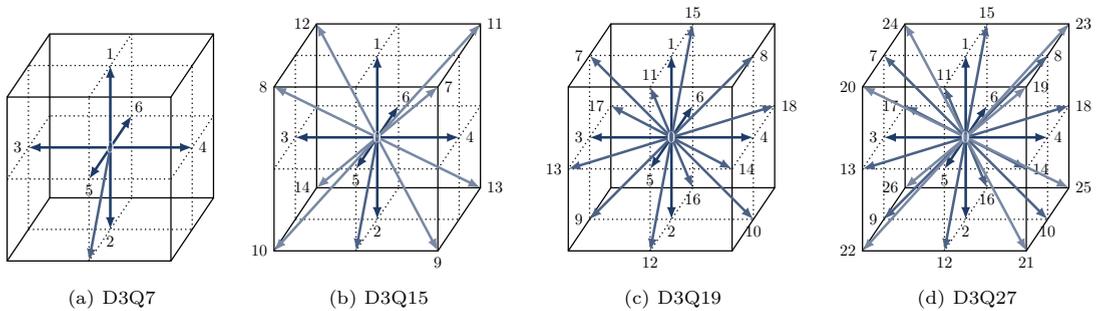
The corresponding weight coefficients are:
\begin{small}
\begin{align}
	w_{q, \mathrm{D3Q7}} &= \frac{1}{6}\left\{
	\begin{array}{l l l}
		& 0, \qquad &q \in \left\lbrace 0 \right\rbrace  \\ 
		& 1, \hfill &q \in \left\lbrace 1, \dots, 6 \right\rbrace \\
	\end{array}\right.\\
w_{q, \mathrm{D3Q15}} &= \frac{1}{72}\left\{
\begin{array}{l l l}
	& 16, \qquad &q \in \left\lbrace 0 \right\rbrace  \\ 
	& 8, \hfill &q \in \left\lbrace 1, \dots, 6 \right\rbrace \\
	& 1, \hfill &q \in \left\lbrace 7, \dots, 14 \right\rbrace
\end{array}\right.\\
w_{q, \mathrm{D3Q19}} &= \frac{1}{36}\left\{
\begin{array}{l l l}
	& 12, \qquad &q \in \left\lbrace 0 \right\rbrace  \\ 
	& 2, \hfill &q \in \left\lbrace 1, \dots, 6 \right\rbrace \\
	& 1, \hfill &q \in \left\lbrace 7, \dots, 18 \right\rbrace
\end{array}\right.\\
w_{q, \mathrm{D3Q27}} &= \frac{1}{216}\left\{
\begin{array}{l l l}
	& 64, \qquad &q \in \left\lbrace 0 \right\rbrace  \\ 
	& 16, \hfill &q \in \left\lbrace 1, \dots, 6 \right\rbrace \\
	& 4, \hfill &q \in \left\lbrace 7, \dots, 18 \right\rbrace\\
	& 1, \hfill &q \in \left\lbrace 19, \dots, 26 \right\rbrace
\end{array}\right.
\end{align}
\end{small}
\section{Droplet motion in three dimensions}
\label{sec:AppendixDroplet3D}

\begin{figure}[htb]
	\centering
	\input{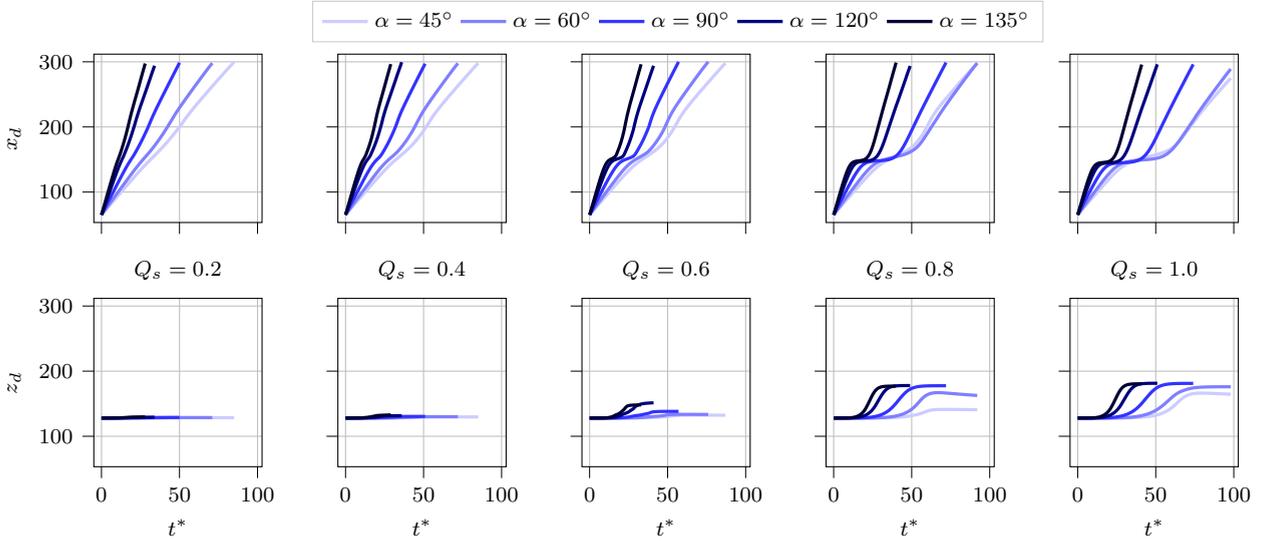}
	\caption{Droplet migration in a three-dimensional channel flow with various contact angles with a single laser point. From left to right, the strength of the laser point is increased by $Q_s = 0.2$, $Q_s = 0.4$, $Q_s = 0.6$, $Q_s = 0.8$, $Q_s = 1$. The figures on the top row show the displacement of the droplets in the $xy$-plane while the bottom row shows the displacement in the $xz$-plane.}
	\label{fig:singleHeatSource}
\end{figure}

Additional simulation results supporting the findings of \cref{sec:3d-laser} are presented. Herein, \cref{fig:singleHeatSource} completes the investigation on three-dimensional droplets using a single heat source.

\begin{figure}[htb]
	\centering
	\input{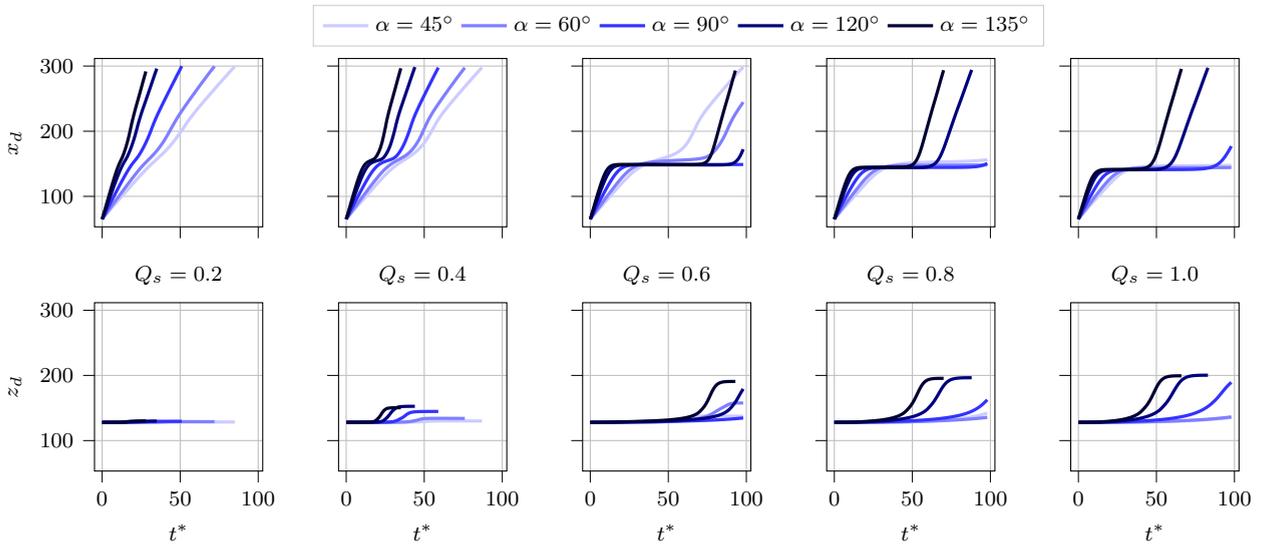}
	\caption{Droplet migration in a three-dimensional channel flow with various contact angles and two single laser point. The two laser points are shifted by $\pm \nicefrac{1}{2}R$ in the $z$-position  From left to right the strength of the laser points is increased by $Q_s = 0.2$, $Q_s = 0.4$, $Q_s = 0.6$, $Q_s = 0.8$, $Q_s = 1$. The figures on the top row show the displacement of the droplets in the $xy$-plane while the bottom row shows the displacement in the $xz$-plane.}
	\label{fig:doubleHeatSource1}
\end{figure}

In \cref{fig:doubleHeatSource1} the investigation is continued by placing a second heat source in the channel. Both heat sources are shifted by $\pm \nicefrac{1}{2}R$ in the $z$-direction. This attempts to show blocking behaviour as in the two-dimensional simulation.

In \cref{fig:doubleHeatSource2} the two heat sources are shifted further apart because in the previous investigation blocking behaviour could not be seen reliable. Here, with a laser strength of $Q_s = 1$ it was possible to block of droplets with contact angles to the bottom wall ranging from $\alpha = 45^{\circ}$ to $\alpha = 135^{\circ}$. The findings demonstrate the additional complexity when transferring two-dimensional thermocapillary flows to three dimensions.

\begin{figure}[htb]
	\centering
	\input{figures/droplet_migration/3D/doubleHeatSource2.tex}
	\caption{Droplet migration in a three-dimensional channel flow with various contact angles and two single laser point. The two laser points are shifted by $\pm \nicefrac{2}{3}R$ in the $z$-position  From left to right the strength of the laser points is increased by $Q_s = 0.2$, $Q_s = 0.4$, $Q_s = 0.6$, $Q_s = 0.8$, $Q_s = 1$. The figures on the top row show the displacement of the droplets in the $xy$-plane while the bottom row shows the displacement in the $xz$-plane.}
	\label{fig:doubleHeatSource2}
\end{figure}
\FloatBarrier

\section*{Acknowledgments}
This work was supported by the SCALABLE project. This project has received funding from the European High-Performance Computing Joint Undertaking (JU) under grant agreement No 956000. The JU receives support from the European Union’s Horizon 2020 research and innovation programme and France, Germany, and the Czech Republic. The authors gratefully acknowledge the Gauss Centre for Supercomputing e.V. (\url{www.gauss-centre.eu}) for funding this project by providing computing time through the John von Neumann Institute for Computing (NIC) on the GCS Supercomputer JUWELS at Jülich Supercomputing Centre (JSC). We acknowledge the EuroHPC Joint Undertaking for awarding this project access to the EuroHPC supercomputer LUMI, hosted by CSC (Finland) and the LUMI consortium through a EuroHPC Regular Access call. The authors gratefully acknowledge the scientific support and HPC resources provided by the Erlangen National High Performance Computing Center (NHR@FAU) of the Friedrich-Alexander-Universität Erlangen-Nürnberg (FAU). TM would like to note that this research was undertaken with the assistance of resources and services from the National Computational Infrastructure (NCI), which is supported by the Australian Government. TM also acknowledges that the work was supported by resources provided by the Pawsey Supercomputing Centre with funding from the Australian Government and the Government of Western Australia.

\section*{Data availability statement}
The complete derivation of methods can be found in the open-source framework \lbmpy{}. For this work, \lbmpy{} version 1.3.3 was used. Two Python notebooks have been added to \lbmpy{} as tutorials to showing the 2D planar heated channel and the 2D droplet dynamics \url{https://pycodegen.pages.i10git.cs.fau.de/lbmpy/}.
The three-dimensional extensions can be found in the \walberla{} source code \url{https://i10git.cs.fau.de/walberla/walberla/-/tree/master/apps/showcases/Thermocapillary}.

\bibliographystyle{elsarticle-num}
\bibliography{2021_TPFLBM_bib.bib}

\end{document}